# A FIRST OPTION CALIBRATION OF THE GARCH DIFFUSION MODEL BY A PDE METHOD

Yiannis A. Papadopoulos[1] and Alan L. Lewis[2]

ABSTRACT

Time-series calibrations often suggest that the GARCH diffusion model could also be a suitable candidate for option (risk-neutral) calibration. But unlike the popular Heston model, it lacks a fast, semi-analytic solution for the pricing of vanilla options, perhaps the main reason why it is not used in this way. In this paper we show how an efficient finite difference-based PDE solver can effectively replace analytical solutions, enabling accurate option calibrations in less than a minute. The proposed pricing engine is shown to be robust under a wide range of model parameters and combines smoothly with black-box optimizers. We use this approach to produce a first PDE calibration of the GARCH diffusion model to SPX options and present some benchmark results for future reference.

# 1. Introduction

Stochastic volatility models are a natural generalization of the seminal Black-Scholes-Merton (BSM) option theory. In such models, the constant volatility parameter $\sigma$ of the BSM theory is promoted to a random process: $dS_t = rS_t\, dt + \sigma_t S_t dW_t^S$. Indeed, there is general agreement in finance that volatility (in its many forms) is best modelled as some sort of mean-reverting stochastic process. Starting from that premise, there are many possibilities. One of the simplest has the instantaneous variance rate $v_t \equiv \sigma_t^2$ evolving as a positive diffusion process following the SDE: $dv_t = \kappa(\bar{v} - v_t)dt + \xi v_t dW_t^v$. Here $W_t^v$ is an additional Brownian motion, $(\kappa, \bar{v}, \xi) > 0$ are constant parameters, and the two Brownian motions $(W_t^S, W_t^v)$ are correlated with constant parameter $\rho$. Coupled with the (risk-neutral) stock price evolution above, this defines the GARCH diffusion model.

The GARCH diffusion model has several nice properties. First, ignoring the drift term for a moment, $v_t$ evolves as a geometric Brownian motion (GBM) -- a natural way in finance to achieve a positive stochastic process. GBM was originally introduced into finance by M. F. M. Osborne in the 1950's to model stock prices under constant volatility. Indeed, time series analysis seems to favor GBM volatility over the popular Heston '93 (square-root) volatility process. Second, with $\bar{v} = 0$, the model nests a variant of the SABR model – very popular in interest rate modelling. The virtue of the SABR-GARCH connection is very tractable small-time behavior – due to a close connection of the small-time dynamics with hyperbolic Brownian motion. (While tractable small-time behavior facilitates time-series analysis by Maximum Likelihood, we found it not especially helpful in option chain calibration). Finally, the model name comes from the property (due to D. Nelson) that there exists a continuous-time limit of a discrete-time GARCH model (GJR-GARCH) that leads to a GARCH diffusion model.[3]

How well can the model fit option chains? Answering that is called calibration. Unfortunately, one desirable – but absent – property is an analytic solution, leaving numerics. While a simulation-based (Monte Carlo) approach doesn't seem like the most efficient approach, in fact the model has been calibrated using Monte Carlo to a large options data set extending over several years by Christoffersen et al. in [1]. They find the GARCH diffusion a better fit than the oft-calibrated Heston '93 model and the so-called 3/2-model, their points of comparison.

[1] Thessaloniki, Greece; email: yianpap99@gmail.com
[2] Newport Beach, California, USA; email: alewis@financepress.com

[3] Briefly summarized (with $\rho = 0$) in Bollerslev and Rossi's 1995 D. Nelson remembrance piece [20].

Given the nice properties, prior calibration results, and the general challenge, we were motivated to develop an efficient, accurate PDE calibrator for this model.[4] Here, we report our methods and first results.

## 1.1 Stock price level-independence (or MAP) property and KBE's.

An important property that the model shares with a wide class of models is stock price level-independence, a well-known scaling relation for vanilla option prices. Specifically, at some initial time $t_0$, consider a vanilla European call option price $C(t_0, S_0, v_0; K, T)$ with strike price $K$, expiration $T$, and state variables $(S_0, v_0)$. Then $C(t_0, S_0, v_0; K, T) = K\, c(t_0, s_0, v_0; T)$ where the standardized option pricing function $c(t, s, v; T)$ is independent of $K$ and $s_0 = S_0/K$. Fixing and suppressing $(K, T)$, consider the pricing function $C(t, S, v)$. It satisfies the KBE (Kolmogorov backwards equation) problem: $-\partial C/\partial t = L_{S,v} C$ with terminal condition $C(T, S, v) = (S - K)^+$, and where $L_{S,v}$ is the process generator. Then, of course, $c(t, s, v)$ satisfies the same PDE with $c(T, s, v) = (s - 1)^+$.

Now fix $K$, say $K = K_0 \equiv S_0$, and solve the (continuum) KBE problem once for expiration $T$. This gives $c(t_0, s, v_0)$, a function of $s$ for $s \in (0, \infty)$ since $c(t_0, s, v_0) = C(t_0, s K_0, v_0)/K_0$ and the r.h.s is known for all values of $s$. For any other strike then, say $K = K_1$, one immediately gets $C(t_0, S_0, v_0; K_1, T) = K_1 c(t_0, S_0/K_1, v_0)$. The point is that a *single* KBE solution yields *all* the (vanilla) option values for different strikes at a given expiration.[5] While obvious in hindsight, the KBE implication of the MAP property initially eluded us. Early on we thought a forward equation (Fokker-Planck) was the only way for pricing "all-options-at-once" at a fixed expiration.[6]

Exploiting the scaling property resulted in significant performance improvements over our original "one option at-a-time" approach: $3\times - 6\times$. Note the improvement ratio is high, but less than the naïve ratio: $N_{options}/N_{expirations}$. The somewhat subtle reasons are discussed in Sec. 3.1.

## 1.2 Our PDE solver in brief.

Given our KBE approach, one must make choices on how to solve the pricing PDE. As with the Heston model, option prices under the GARCH diffusion model are governed by a 2-D convection-diffusion-reaction PDE with a mixed derivative term. Key characteristics of a suitable numerical scheme would be a) stability under practical usage, b) good accuracy to execution time ratio, and c) robustness (good oscillation-damping properties). As noted in [2], spurious oscillations in numerical computation of option prices can have three distinct causes: convection dominance, time-stepping schemes that are unable to sufficiently damp the high frequency errors stemming from the payoff discontinuity, and finally negative coefficients arising from the discretization of the diffusion terms. Here we take a closer look at the last two.

For the spatial discretization we use the finite difference method on non-uniform grids. We employ standard central finite difference formulae for the diffusion and convection terms, but opt for a less common formula for the mixed derivative term, one that helps reduce oscillations that may take the solution to negative values. Although not our first choice, we also discretize the PDE cast in the natural logarithm of the asset price; combined with the mixed derivative scheme, this can further guard against negative values (but not preclude them altogether).

---

[4] We are well-aware of the general limitations of simple stochastic volatility diffusions. For example, they have difficulty fitting short-dated SPX option smiles and VIX options. Overcoming the limitations seems to require jump-processes. But, even if you want to include jumps into so-called "non-affine" models (like the GARCH diffusion), you need to start with a good PDE solver.

[5] For American options, barrier options, and other more exotic options, individual KBE solutions are needed.

[6] More generally, replace $v_0$ in the scaling argument above by $Y_0$, a (D-1) vector-valued state variable for a D-dimensional jump-diffusion or whatever. Then, scaling (and thus "all-options-at-once") for Euro-style vanillas holds if: the process $(X_t, Y_t)$ is a MAP (Markov Additive Process), where $X_t \equiv \log S_t$ is the additive component and $Y_t$ is the Markov component. MAPs are defined in Çinlar [19]; modelling implications are stressed in Lewis (2016) [5]. Note this generality admits even discrete-time processes. Thus, for MAPs, it suffices to solve the backwards evolution problem once for a single strike to get all the vanilla option prices at a given expiration $T$. Admitting jumps, the backwards evolution problem (in continuous-time) is generally a PIDE (partial integro-differential equation) problem.

With spatial discretization in place, one is left with a large system of stiff ordinary differential equations and must adopt a time-marching method. We employ two commonly used schemes plus a rather unusual one. For this type of PDE the most popular choice would be a cross-derivative-enabled ADI variant (see [3] for an overview). We opt for the Hundsdorfer-Verwer and the Modified Craig-Sneyd schemes that offer the best overall characteristics. Our alternative is the BDF3 fully implicit scheme which, as far as we know, has not been used in such a context in the financial literature.

It may have already become apparent that we do not aim for one sole scheme that is necessarily monotone by design; we believe that such a scheme would likely be less accurate or slower than it needs to be. What we aim for instead is a reliable set-up, enabling as fast and accurate calibrations as possible. To this end we propose a strategy that involves occasional re-evaluations and a hybrid engine that switches from ADI to the slower but more robust BDF3 scheme in such cases.

The optimization is done with commercial software. We mainly use local constrained optimization routines, but we also try a global method (Differential Evolution). The latter, while proving too slow to be the recommended option, can be used to add confidence that the local optimizer is indeed finding global minima (which we have found to be the case in all our tests).

The rest of this paper is organized as follows: Sec. 2 presents the numerical methods for the solution of the pricing PDE, with non-standard implementation specifics given in more detail. Sec. 3 describes the calibration phase and proposed strategy for optimizing performance. Sec. 4 contains various numerical results. We compare the computational efficiency of the time-marching schemes and examine the effectiveness of Richardson extrapolation in both space and time. This is followed by reference calibration results to real data and comparisons with the Heston model. We conclude with a brief exploration of other non-affine models that are readily handled by our framework. We finally present our conclusions and suggestions for further development.

## 2. Numerical solution of the GARCH diffusion PDE

### 2.1. The GARCH diffusion PDE.

The GARCH diffusion stochastic volatility model is described (under the risk-neutral measure) by

$$dS_t = (r_T - q_T)S_t dt + \sqrt{v_t} S_t dW_t^S,$$
$$dv_t = \kappa(\bar{v} - v_t)dt + \xi v_t dW_t^v. \tag{1}$$

Here the Brownian noises associated to the underlying asset $S_t$ (here the SPX) and its variance $v_t$ are correlated; i.e., $dW_t^S dW_t^v = \rho\, dt$. A compatible real-world evolution is given in Appendix A. Time-series analysis (of similar real-world models) suggests that the correlation coefficient $\rho$ is negative with typical values of around -0.75 (Ait-Sahalia & Kimmel [4]). So here we assume $\rho < 0$. The variance process $v_t$ has volatility $\xi > 0$ and reverts to its long-run mean $\bar{v} > 0$ with a mean-reversion rate of $\kappa > 0$. $T$ is the time of an option expiration. Generally, our model assumes an environment with deterministic interest rate and dividend yields: $(r_t, q_t)$. But we write $(r_T, q_T)$ to indicate that we are using stepwise constants for each option expiration. (There will be some deterministic behavior for $(r_t, q_t)$ compatible with this).

Let then $V(S, v, t)$ denote the price of a European option when at time $T - t$ the underlying asset price equals $S$ and its variance equals $v$. It is easy to verify that under the above specification $V(S, v, t)$ must satisfy the following parabolic PDE

$$\frac{\partial V}{\partial t} = \frac{1}{2}S^2 v \frac{\partial^2 V}{\partial S^2} + \rho\xi S v^{3/2} \frac{\partial^2 V}{\partial S \partial v} + \frac{1}{2}\xi^2 v^2 \frac{\partial^2 V}{\partial v^2} + (r_T - q_T)S\frac{\partial V}{\partial S} + \kappa(\bar{v} - v)\frac{\partial V}{\partial v} - r_T V \tag{2}$$

for $0 \leq t \leq T$, $S > 0$, $v > 0$. We can also cast the equation in terms of the natural logarithm of the price $X = \ln(S)$

$$\frac{\partial V}{\partial t} = \frac{1}{2}v\frac{\partial^2 V}{\partial X^2} + \rho\xi v^{3/2}\frac{\partial^2 V}{\partial X \partial v} + \frac{1}{2}\xi^2 v^2 \frac{\partial^2 V}{\partial v^2} + \left(r_T - q_T - \frac{1}{2}v\right)\frac{\partial V}{\partial X} + \kappa(\bar{v} - v)\frac{\partial V}{\partial v} - r_T V \quad (3)$$

for $0 \leq t \leq T$, $v > 0$. Equations (2) and (3) are categorized as time-dependent convection-diffusion-reaction PDE's on an unbounded spatial domain. While (3) has a slightly simpler form, it is harder to allocate points on the $X$-grid optimally. This is especially true since in many cases the grid needs to start from very small $S$ values (to avoid loss of accuracy from grid truncation), which then means that a lot of $X$-points will be placed in an area of low interest. We will thus discretize and solve (2) primarily, but that the code is also (trivially) adapted for switching to solving (3) as well.

**Initial and boundary conditions**

As initial conditions to (2) we have the vanilla call and put payoffs

$$V_{call}(S, v, 0) = \max(S - K, 0), \quad V_{put}(S, v, 0) = \max(K - S, 0), \quad (4)$$

where K is the strike of the option. We impose (numerical) boundary conditions of Dirichlet type (5) - (6) and Neumann type (7) - (9) on the left and right-side boundaries respectively:

$$V_{call}(S_{min}, v, t) = 0, \quad (5)$$

$$V_{put}(S_{min}, v, t) = Ke^{-r_T t} - S_{min} e^{-q_T t}, \quad (6)$$

$$\frac{\partial V_{call}}{\partial S}(S_{max}, v, t) = e^{-q_T t}, \quad (7)$$

$$\frac{\partial V_{put}}{\partial S}(S_{max}, v, t) = 0, \quad (8)$$

$$\frac{\partial V_{call}}{\partial v}(S, v_{max}, t) = \frac{\partial V_{put}}{\partial v}(S, v_{max}, t) = 0. \quad (9)$$

Under this model $v = 0$ is an entrance boundary for all $(\kappa, \xi) > 0$, meaning that $v = 0$ is unreachable whenever the process starts at $v_o > 0$. However, the process may in principle be started at $v_o = 0$, after which it immediately enters the interior and never hits the origin again[7] (for more details the reader is referred to Lewis [5], pg. 102). Therefore, from a mathematical standpoint no boundary condition is necessary. The PDE itself can be applied at $v = v_{min} = 0$ (where all diffusion terms vanish due to the presence of factor $v$) and there is no need for any extra condition from a numerical point of view either. The choice of the grid truncation boundaries $S_{min}, S_{max}$ (or $X_{min}, X_{max}$) and $v_{max}$ is discussed in Sec. 2.4.1.1. The boundary conditions are set in an equivalent manner for equation (3).

## 2.2. Spatial discretization

We discretize in space using the finite difference method and work on non-uniform grids which we consider necessary for the efficient solution of the pricing PDE. In the $S$-direction, allocating more points around the strike can significantly reduce the error stemming from the initial delta discontinuity there. In the $v$-direction, allocating more points near $v_0$ makes sense since we want to resolve better the area where we want to obtain a price. Also, since typically we have $v_{max} \gg v_o$ (see Sec. 2.4.1.1), a non-uniform $v$-grid is all but necessary to both adequately resolve the area around $v_o$ and at the same time reach out to $v_{max}$ with a reasonable number of grid points.

We use the standard central finite difference formulas for the first and second derivatives in (2) and (3) and a rather less standard seven-point stencil representation for the mixed derivative. All formulas give second-order accurate approximations, provided the grid step variation is sufficiently smooth (as is indeed the case for the grid construction proposed in Sec. 2.4.1.2).

---

[7] This means that there are indeed non-trivial option price solutions for $v_0 = 0$.

Let the grid in the $S$-direction be defined by $NS + 1$ points, $0 \leq S_{min} = S_0 < S_1 < \cdots < S_{NS} = S_{max}$ and the corresponding grid steps $\Delta S_i = S_i - S_{i-1}$, $i = 1, 2, \ldots, NS$. We then define the discretized versions of the first and second derivatives $\partial V_{i,j}/\partial S$ and $\partial^2 V_{i,j}/\partial S^2$ at $S = S_i$ as

$$\frac{\partial V_{i,j}}{\partial S} \approx \frac{-\Delta S_{i+1}}{\Delta S_i (\Delta S_i + \Delta S_{i+1})} V_{i-1,j} + \frac{\Delta S_{i+1} - \Delta S_i}{\Delta S_i \Delta S_{i+1}} V_{i,j} + \frac{\Delta S_i}{\Delta S_{i+1}(\Delta S_i + \Delta S_{i+1})} V_{i+1,j}, \quad (10)$$

$$\frac{\partial^2 V_{i,j}}{\partial S^2} \approx \frac{2}{\Delta S_i (\Delta S_i + \Delta S_{i+1})} V_{i-1,j} - \frac{2}{\Delta S_i \Delta S_{i+1}} V_{i,j} + \frac{2}{\Delta S_{i+1}(\Delta S_i + \Delta S_{i+1})} V_{i+1,j}. \quad (11)$$

We use the equivalent expressions for the derivatives $\partial V_{i,j}/\partial v$ and $\partial^2 V_{i,j}/\partial v^2$ at $v = v_j$ in the $v$-direction where the grid is defined by $NV + 1$ points, $0 = v_0 < v_1 < \cdots < v_{Nv} = v_{max}$ and $\Delta v_j = v_j - v_{j-1}$, $j = 1, 2, \ldots, NV$. An exception is the $v = 0$ boundary where we use the one-sided (upwind) second-order formula[8] for $\partial V_{i,j=0}/\partial v$:

$$\frac{\partial V_{i,0}}{\partial v} \approx -\frac{(2\Delta v_1 + \Delta v_2)}{\Delta v_1 (\Delta v_1 + \Delta v_2)} V_{i,0} + \frac{(\Delta v_1 + \Delta v_2)}{\Delta v_1 \Delta v_2} V_{i,1} - \frac{\Delta v_1}{\Delta v_2 (\Delta v_1 + \Delta v_2)} V_{i,2}. \quad (12)$$

For the mixed derivative term, we opt for a custom second-order scheme based on a 7-point stencil, which is very similar but not identical to that proposed by Ikonen & Toivanen in [6]. Such a scheme can be constructed so that it contributes fewer negative off-diagonal coefficients to the resulting system's discretization matrix **A** than the standard second-order scheme based on the 9-point stencil. This in turn makes the solution less likely to produce a negative valuation. When the correlation coefficient $\rho$ is negative (which we take to be the case here as discussed in Sec. 2.1), an appropriate formula for approximating $\partial^2 V_{i,j}/\partial S \partial v$ at $(S, v) = (S_i, v_j)$ is given by

$$\frac{\partial^2 V_{i,j}}{\partial S \partial v} = \frac{1}{D}\left(-V_{i+1,j-1} + 2V_{i,j} - V_{i-1,j+1} + (\Delta S_{i+1} - \Delta S_i)\frac{\partial V_{i,j}}{\partial S} + (\Delta v_{j+1} - \Delta v_j)\frac{\partial V_{i,j}}{\partial v} + \frac{1}{2}(\Delta S_{i+1}^2 + \Delta S_i^2)\frac{\partial^2 V_{i,j}}{\partial S^2} + \frac{1}{2}(\Delta v_{j+1}^2 + \Delta v_j^2)\frac{\partial^2 V_{i,j}}{\partial v^2}\right) \quad (13)$$

where
$$D = \Delta S_{i+1} \Delta v_j + \Delta S_i \Delta v_{j+1}. \quad (14)$$

Formula (13) is readily obtained considering Taylor expansions of the option value $V_{i,j}$ at the neighboring upper left and lower right grid points $(S_{i-1}, v_{j+1})$ and $(S_{i+1}, v_{j-1})$. Such a formula can be used in conjunction with a specially constructed grid (with limitations imposed on the grid steps) and some use of first-order upwind formulas for the convection terms $\partial V_{i,j}/\partial S$ and $\partial V_{i,j}/\partial v$, to make **A** an M-matrix by design (see [6] for example). This would ensure that the solution cannot produce a negative valuation in any case, which would be a particularly useful feature for our calibration (that requires the calculation of implied volatilities). Such an approach though is not favored in the present work, since we believe that it would unnecessarily reduce the average accuracy of the solution through suboptimal grid construction: The grid points allocation should be driven by the problem's physical characteristics (e.g. the location of the payoff discontinuity) and not be forced upon through the mathematical requirement of nonnegative coefficients[9]. We revisit this in Sec. 3 where we explain how we handle the occasional negative values that are indeed possible under the proposed discretization.

We can now replace the spatial derivatives on the right-hand side of equation (2) with their discretized versions described above to obtain its semi-discretized form

---

[8] Using the first-order (two-point) upwind formula would be better from a stability point of view, but would result in loss of accuracy and the overall second-order convergence of the discretization. In practice we have seen no stability issues arising from the use of (12) in extensive tests throughout numerous calibration exercises.

[9] Zvan et al. [2] provide a similar discussion, albeit in the context of finite volume/element discretization.

$$\frac{\partial V_{i,j}}{\partial t} = d_S \frac{\partial^2 V_{i,j}}{\partial S^2} + d_v \frac{\partial^2 V_{i,j}}{\partial v^2} + c_S \frac{\partial V_{i,j}}{\partial S} + c_v \frac{\partial V_{i,j}}{\partial v} + m_{Sv}(-V_{i+1,j-1} + 2V_{i,j} - V_{i-1,j+1}) - r_T V_{i,j}, \quad (15)$$

where the diffusion, convection and mixed derivative coefficients are given by

$$d_S = \frac{1}{2} S_i^2 v_j + \frac{1}{2} m_{Sv}(\Delta S_{i+1}^2 + \Delta S_i^2), \quad (16)$$

$$d_v = \frac{1}{2} \xi^2 v_j^2 + \frac{1}{2} m_{Sv}(\Delta v_{j+1}^2 + \Delta v_j^2), \quad (17)$$

$$c_S = (r_T - q_T) S_i + m_{Sv}(\Delta S_{i+1} - \Delta S_i), \quad (18)$$

$$c_v = \kappa(\bar{v} - v_j) + m_{Sv}(\Delta v_{j+1} - \Delta v_j), \quad (19)$$

$$\text{and } m_{Sv} = \frac{1}{D} \rho \xi S_i v_j^{3/2}. \quad (20)$$

Equation (15) is applied at each grid point $(S_i, v_j)$ for $i = 1, 2, \ldots NS$ and $j = 0, 1, \ldots, NV$. We do not need to solve for $S_{i=0} = S_{min}$ since the Dirichlet boundary conditions (5) and (6) specify constant values there for the option value $V$. At $v_{j=0} = 0$ the second and mixed derivative terms in (15) vanish and the upwind discretization (12) means we only use values within our grid. This is not the case though for the far-boundary grid lines $S_{i=NS} = S_{max}$ and $v_{j=NV} = v_{max}$: the Neumann-type conditions (7) - (9) imply that the mixed derivative $\partial^2 V / \partial S \partial v$ (and thus all the terms in (15) - (19) multiplied by $m_{Sv}$) vanish. But we still have the second derivatives, whose stencils reference a point outside the grid. Such points are treated as fictitious and their value is obtained through extrapolation based on the last actual grid point and the known value of the gradient there.

### 2.3. Time discretization

With spatial discretization in place we are now left with a large system of stiff ordinary differential equations (ODE's) in time, which we can write as

$$\boldsymbol{V}'(t) = \boldsymbol{F}(t, \boldsymbol{V}(t)), \ \boldsymbol{V}(0) = \boldsymbol{V_0}, \quad \text{where } \boldsymbol{F}(t, \boldsymbol{V}(t)) := \boldsymbol{A}\boldsymbol{V}(t) + \boldsymbol{b}(t) \text{ for } 0 \le t \le T. \quad (21)$$

Here $\boldsymbol{V}'(t), \boldsymbol{V}(t), \boldsymbol{b}(t)$ and $\boldsymbol{V_0}$ are vectors of size $M$ and $\boldsymbol{A}$ is the $M \times M$ spatial discretization matrix, where $M = NS \times (NV + 1)$ is the total number of unknowns. The elements of $\boldsymbol{b}(t)$ will depend on the boundary conditions (5) - (9) and those of $\boldsymbol{V_0}$ on the initial conditions (4).

We now need to adopt a time-marching method to solve (21). Popular choices for 1-D problems, such as the Implicit Euler and Crank-Nicolson schemes, become inefficient in higher dimensions, leading to large sparse systems that are a lot more expensive to solve than the small (typically tridiagonal) ones in the 1-D case. ADI-type splitting schemes are thus the most popular choice in 2-D and 3-D. However, standard (non-splitting) schemes can still be competitive for 2-D problems if a fast, sparse direct solver is used. This is especially true for the vanilla option pricing problem as the coefficients are time-independent and the matrix factorization step only needs to be performed once (or a few times). We employ one such method not often used in finance, namely the BDF3 (or 4-Level Fully Implicit) scheme. Our main workhorses though will be two popular ADI schemes which we briefly present first.

### 2.3.1 ADI schemes

For a detailed review of ADI methods for PDE's with mixed derivatives in finance, the reader is referred to [3]. The first step for all such methods is to decompose $\boldsymbol{A}$ in (21) into three submatrices:

$$\boldsymbol{A} = \boldsymbol{A_0} + \boldsymbol{A_1} + \boldsymbol{A_2}. \quad (22)$$

$A_0$ contains all terms stemming from the discretization of the mixed derivative term in (2), (3), i.e., all terms in (15) including $m_{Sv}$ as a factor. $A_1$ and $A_2$ contain all the terms corresponding to the discretized derivatives in the *S*-direction and *v*-direction respectively. The source term $r_T V$ is evenly distributed between $A_1$ and $A_2$. By virtue of our 3-point central discretizations for the convection and diffusion terms, $A_1$ and $A_2$ are tridiagonal matrices[10]. We split the vector $b(t)$ and function $F(t, V)$ from (21) accordingly as $b(t) = b_0(t) + b_1(t) + b_2(t)$ and $F(t, V) = F_0(t, V) + F_1(t, V) + F_2(t, V)$. We will use a uniform temporal grid which is defined by the points $t_n = n \cdot \Delta t,\ 0 \leq n \leq NT,\ \Delta t = \frac{T}{NT}$. Let $\theta$ be a real parameter which will control the exact splitting.

We now outline our two main schemes, chosen for their optimal combination of stability, accuracy and inherent oscillation-damping properties [7].

*Hundsdorfer-Verwer (HV) scheme*

$$\begin{cases} Y_0 = V_{n-1} + \Delta t F(t_{n-1}, V_{n-1}), & \text{step 1} \\ Y_k = Y_{k-1} + \theta \Delta t \left( F_k(t_n, Y_k) - F_k(t_{n-1}, V_{n-1}) \right)\ (k=1,2), & \text{steps 2 \& 3} \\ \tilde{Y}_0 = Y_0 + {}^1\!/_2 \Delta t \left( F(t_n, Y_2) - F(t_{n-1}, V_{n-1}) \right), & \text{step 4} \\ \tilde{Y}_k = \tilde{Y}_{k-1} + \theta \Delta t \left( F_k(t_n, \tilde{Y}_k) - F_k(t_n, Y_2) \right)\ (k=1,2), & \text{steps 5 \& 6} \\ V_n = \tilde{Y}_2 \end{cases}$$

*Modified Craig-Sneyd (MCS) scheme*

$$\begin{cases} Y_0 = V_{n-1} + \Delta t F(t_{n-1}, V_{n-1}), & \text{step 1} \\ Y_k = Y_{k-1} + \theta \Delta t \left( F_k(t_n, Y_k) - F_k(t_{n-1}, V_{n-1}) \right)\ (k=1,2), & \text{steps 2 \& 3} \\ \hat{Y}_0 = Y_0 + \theta \Delta t \left( F_0(t_n, Y_2) - F_0(t_{n-1}, V_{n-1}) \right), & \text{step 4} \\ \tilde{Y}_0 = \hat{Y}_0 + ({}^1\!/_2 - \theta) \Delta t \left( F(t_n, Y_2) - F(t_{n-1}, V_{n-1}) \right), & \text{step 5} \\ \tilde{Y}_k = \tilde{Y}_{k-1} + \theta \Delta t \left( F_k(t_n, \tilde{Y}_k) - F_k(t_{n-1}, V_{n-1}) \right)\ (k=1,2), & \text{steps 6 \& 7} \\ V_n = \tilde{Y}_2 \end{cases}$$

Both schemes employ multiple intermediate steps to advance the solution from $V_{n-1}$ to $V_n$. The HV scheme starts with a forward Euler (predictor) step (1), followed by two unidirectional implicit (corrector) steps (2 & 3) which serve to stabilize the explicit first step. Then a second predictor step (4) is followed by two more implicit corrector steps (5 & 6). The MCS scheme has an identical structure except for the double second predictor step (steps 4 & 5). The implicit steps require the solution of tridiagonal systems which we solve efficiently with LU decomposition. We use the HV scheme with $\theta = 1 - \sqrt{2}/2$ (which we shall refer to as HV1) and $\theta = 1/2 + \sqrt{3}/6$ (HV2). It was conjectured in [3] that HV1 is only conditionally stable (but more accurate), and HV2 unconditionally stable (and less accurate). For the MCS scheme we use $\theta = 1/3$, recommended in [8] as an optimal value based on stability analysis and experiments. Regardless of the value of $\theta$, both schemes are second-order.

We note that despite proven unconditionally (von Neumann-) stable, these schemes do not always sufficiently damp local high-frequency errors caused by discontinuities in the initial conditions. This may result in spurious oscillations and reduced order of convergence; see for example [3, 9]. In this case a technique known as Rannacher time-stepping can be used to palliate the issue. This involves using a different scheme for the first time-step (which is divided into two equal sub-steps), one that can successfully damp oscillations and is usually first-order (typically the Euler Implicit scheme).

---

[10] This is not strictly true for matrix $A_2$ because the one-sided formula (12) used for the $v = 0$ boundary involves one more point off the diagonal.

### 2.3.2 The BDF3 scheme

For a more robust alternative that could conceivably provide smoother inputs to the (gradient-based) optimizer, we look at the third-order BDF3 scheme. Although not a typical choice, it is nonetheless simple to implement and has good stability (it is almost A-stable in an ODE sense) and oscillation damping properties. To solve the resulting systems, we use the Eigen C++ matrix library that offers simple interfaces to several direct sparse system solvers. The fastest one for the present system structure seems to be the UMFPACK solver, which we used for our experiments here.

The scheme simply amounts to replacing the time derivative $V'(t)$ in (21) with a one-sided, 4-level backward finite difference expression. The discretized version of (21) then looks like

$$\frac{\frac{11}{6}V_n - 3V_{n-1} + \frac{3}{2}V_{n-2} - \frac{1}{3}V_{n-3}}{\Delta t} = AV_n + b_n = F(t_n, V_n), \tag{25}$$

and the values $V_n$ at time level $n$ are calculated given the values at the previous three time-levels as

$$\frac{11}{6}V_n = 3V_{n-1} - \frac{3}{2}V_{n-2} + \frac{1}{3}V_{n-3} + \Delta t(AV_n + b_n). \tag{26}$$

Since values are required not only from the previous time-level (like the ADI methods), but also from two levels before that, we must use some alternative scheme for the first two steps of the integration. We use the first-order Implicit Euler (IE) scheme and the second-order BDF2 scheme. The IE scheme is given by $V_n = V_{n-1} + \Delta t(AV_n + b_n)$ and requires the factorization of $A_{IE} = (I - \Delta t A)$. The BDF2 scheme is given by $1.5V_n = 2V_{n-1} - 0.5V_{n-2} + \Delta t(AV_n + b_n)$ and requires the factorization of $A_{BDF2} = (1.5I - \Delta t A)$. In order to improve accuracy for the first time-step, we employ Richardson extrapolation like this: we first use the IE scheme for 4 sub-steps of size $\Delta t/4$ to get the values $V_1^{fine}$ at the end of the first time-step. We then repeat, this time using 2 sub-steps of size $\Delta t/2$ to obtain $V_1^{coarse}$ and get the final composite values for the first time-step as $V_1 = 2V_1^{fine} - V_1^{coarse}$. Note that this requires 2 matrix factorizations, corresponding to $A_{IE}$ with $\Delta t/4$ and $\Delta t/2$. To get the values $V_2$ at the end of the second time-step we use the BDF2 scheme. In total, the present implementation requires 4 expensive factorizations which add a substantial upfront computational cost.

### 2.4. Increasing computational efficiency

Let us loosely define computational efficiency (CE) as the accuracy achieved per unit CPU time. A PDE-based solver cannot match the CE of semi-analytical solutions, such as those available for the Heston model. We therefore need to look into ways of improving the CE of our set-up. Here we consider grid construction, smoothing of the initial conditions and Richardson extrapolation.

### 2.4.1 Grid construction

#### 2.4.1.1. Grid truncation

Our domain is semi-infinite (or infinite for X in equation (3)), so in practice the grid needs to be truncated at some point. If the grid does not extend far enough then the imposed boundary conditions will not hold exactly true and forcing them on the solution will introduce some error. If the grid extends further than it needs to then the grid step sizes will be larger for the same number of points, resulting in less accurate finite difference approximations. There is no obvious way to determine the truncation limits, so here we make the empirical choices below. Note the dependence of the limits on the model parameters, which means that the grids used will be different for each objective function evaluation (based on the parameters set by the optimizer each time).

*S-direction*

For the $S$-grid, we truncate to the right at $S_{max} = e^{(ln(max(K,S_0)) + M\sigma_{est}\sqrt{T})}$, where we set M = 5 and $\sigma_{est} = 0.5(\sqrt{v_0} + \sqrt{v_L})$. We then set: $1.5K < S_{max} < 20K$. This choice leads to good solution accuracy overall, but for extreme model parameter regimes and benchmark calculations we additionally multiply $S_{max}$ by a safety factor of 2 to 3. For the left boundary and for equation (2), we set $S_{min} = 0$. For equation (3) in $X = \ln(S)$, we truncate at $X_{min} = \ln(min(K, S_0)) - M\sigma_{est}\sqrt{T}$, where we set M = 6. We then further require that $X_{min} \leq \alpha K$, where $\alpha$ is some constant. We normally set $\alpha = 0.1$ but for high accuracy we recommend $\alpha \leq 0.025$.

*v-direction*

We set $v_{min} = 0$, i.e., we do not truncate the left boundary. To set an appropriate right boundary, we note that for $T \to \infty$, $v_t$ follows an Inverse Gamma distribution (see Appendix B). Given the distribution we can then set

$$v_{max} = v_{crit}(q) = F'(q) \qquad (27)$$

and $F'$ is the inverse cumulative (Inverse Gamma) probability function. We find that a value of $1 - q$ between $1 \cdot 10^{-5}$ and $1 \cdot 10^{-6}$ is necessary for accurate valuations. For short-dated options an empirical fraction of (27) can be used whenever $\kappa \cdot T < 1$. Alternatively, one can numerically calculate the exact distribution – and thus $v_{crit}(q)$ – for each expiration. This is described in Appendix B and is used for our experiments in Sec. 4 with $1 - q = 2.5 \cdot 10^{-6}$. We finally note that typically it will be $v_{max} \gg v_0$. This observation alone necessitates the use of a non-uniform grid, described next.

### 2.4.1.2. Grid generation

Computational efficiency can be improved significantly and any problems due to discontinuities mitigated, with a grid that concentrates more points where they're needed. We employ a well-known one-dimensional grid-generating (stretching) function based on the inverse hyperbolic sine, which satisfies certain criteria for use with finite difference methods. The interested reader is referred to Vinokur [10]. The same function but in slightly different form is often used in the financial literature, see for example Tavella & Randall [11] and In 't Hout & Foulon [3].

The grid in the S-direction is given by:

$$S_i = S_{min} + K\left(1 + \sinh\left(b_S\left(\frac{i}{NS} - a_S\right)\right)/\sinh(b_S a_S)\right) \qquad (28)$$

for $i = 0, 1, \ldots, NS$, where K is the strike (or more generally the desired clustering point) and $a_S, b_S$ are free parameters. $a_S$ represents the percentage of total points that lie between $S_{min}$ and K and $b_S$ controls the degree of non-uniformity. We set $b_S = 4.5$ which corresponds to moderate non-uniformity and generally results in low error profiles across the moneyness spectrum. Given $b_S$, $a_S$ can be set so that the grid goes up to $S_{max}$ using:

$$a_S = \ln\left((A + e^{b_S})/(A + e^{-b_S})\right)/2b_S, \text{ where } A = (S_{max} - K)/(K - S_{min}).$$

We then make sure that the strike falls exactly on a grid point by making a further slight adjustment to $a_S$: we find $i_K = \lfloor a_S NS \rfloor$ and then reset $a_S = i_K/NS$.[11] Finally, we use the same approach for generating the X-grid in equation (3).

For the v-direction we again use the same grid-generating function:

$$v_j = v_0\left(1 + \sinh\left(b_v\left(\frac{j}{NV} - a_v\right)\right)/\sinh(b_v a_v)\right), \qquad (29)$$

for $j = 0, 1, \ldots, NV$, which clusters points around $v_0$. Since $v_{max} \gg v_0$ we set $b_v = 8.5$ which is as non-uniform as we can get before CE starts dropping. We first set $a_v$ so that the grid goes up to $v_{max}$:

$$a_v = \ln\left((A + e^{b_v})/(A + e^{-b_v})\right)/2b_v, \text{ where } A = \frac{v_{max}}{v_0} - 1.$$

---

[11] If we wanted to place the strike in the middle between grid points we would use $a_S = (i_K + 0.5)/NS$ instead.

We then find $N_{v_0} = \max(\lfloor a_v NV \rfloor, \lfloor 0.2NV \rfloor, 6)$ [12] and reset $a_v = N_{v_0}/NV$ to ensure that $v_0$ lies exactly on a grid point. When the input $NV$ is low and/or $v_{max} \gg v_0$, then the above $a_v$ adjustment results in the last grid point now falling short of $v_{max}$. In such cases we keep adding points using (29) until $v_j \geq v_{max}$, which means that the final grid size will be $NV^*$, with $NV^* \geq NV$. Typically, $NV^*$ will be up to 50% higher than the input $NV$.

An alternative construction that seems to have some advantage over the one just described, is a hybrid one, having the narrow (but most important) zone around $v_0$ uniformly-spaced and the rest non-uniform. Haentjens & In 't Hout [12] propose one way of constructing such a grid in the $S$-direction for the solution of the Heston PDE. Here we use our first construction above as the base, to determine $N_{v_0}$ and $NV^*$. The segment $(0, 2v_0)$ is then made uniform with step $\Delta v_U = v_0/N_{v_0}$. We then use the simple stretching function

$$v_j = R_{NU} \sinh\left(\frac{b_v j}{N_{NU}}\right)/\sinh(b_v), \quad N_{NU} = NV^* - 2N_{v_0} + 1, \quad R_{NU} = v_{max} - 2v_0 + \Delta v_U \tag{30}$$

to generate the non-uniform part, choosing $b_v$ so that the first step is equal to $\Delta v_U$. This can be easily achieved with any one-dimensional root-finding method.

The two $v$-grid constructions generally result in comparable performance. But when used with Richardson extrapolation (Sec. 2.4.3), the second (hybrid) variant is always preferable. We thus use the hybrid construction for all the numerical experiments of Sec. 4.

### 2.4.2. Smoothing of the initial conditions

Whenever there is a discontinuity at some point in the initial conditions, it is usually a good idea to apply some sort of averaging for that point using the value(s) of adjacent point(s). That is effectively to smooth out the discontinuity (in this case located at the strike $K$) before solving the PDE. The reason is that such discontinuities increase the solution error. To this end, here we just replace the (zero) initial condition values along the $S = K$ line of the grid (remember we made sure that there is a grid point $S_{i_K}$ on $K$) with a simple average over nearby space as proposed in [13]. For vanilla options this amounts to setting:

$$initCond_{i_K,j} = \frac{0.25\Delta S^2}{S_{i_K+1} - S_{i_K-1}},$$

for $j = 0, 1, \ldots, NV$, where we have $\Delta S = S_{i_K+1} - S_{i_K}$ for calls and $\Delta S = S_{i_K} - S_{i_K-1}$ for puts.

### 2.4.3. Richardson extrapolation (spatial)

Richardson extrapolation (RE) can significantly increase accuracy for many problems adding only a small computational overhead. It simply involves calculating solutions based on two different grids (either spatial or temporal, usually with grid-step sizes ratio of 2:1) and "combining" them based on the discretization's theoretical order of convergence. Here we apply it on the spatial level as follows: for a given resolution $NS \times NV$, we first generate a $(NS/2 \times NV/2)$ grid and calculate an option price on it, $P_{coarse}$. Then using the same grid parameters $(a_S, b_S)$ and $(a_V, b_V)$, we generate a $(NS \times NV)$ grid and use it to calculate $P_{fine}$. Given that our discretization is full second-order in both $S$ and $v$, we can then calculate the extrapolated price as $P_{RE} = \frac{4}{3}P_{fine} + \frac{1}{3}P_{coarse}$. Note that the fine grid will contain all the coarse grid's points and add new ones in between. It is important that the relative location of the strike $K$ is the same for the two grids, as is indeed the case (both have points exactly on the $S = K$ line). The main advantage is that while the computational cost increase is merely 25%, the accuracy is typically improved by 1-2 orders of magnitude (depending on the resolution used and the model parameters).

RE works very well when sufficiently fine grids are used and not so well when the grids are too coarse (in which case it may well give worse accuracy than the single evaluation). This is because the premise for RE is that the two solutions are in the asymptotic range, i.e., that the observed order of convergence for the grids used is (very close to) the theoretical one. Down to the lowest resolution $(NS \times NV) = (40 \times 20)$ used in our experiments, we've found RE to clearly outperform the single evaluation in terms of CE.[13]

---

[12] To guarantee that the solution around $v_0$ is always adequately resolved, we make sure that there is a minimum number of allocated grid points up to $v_0$, at least 20% of the total and no less than 6.

RE is also less effective for 2-D and 3-D problems when non-uniform grids with different stretching functions for each dimension are used (as is the case here). We find that this is effectively countered with the use of the hybrid $v$-grid that makes the grid in the $v$-direction uniform in the region of interest. This helps to regularize convergence, which in turns leads to improved RE performance.

Discontinuities and/or singularities in the initial or boundary conditions will also often cause the observed order of convergence to be less than the theoretical one (and make convergence overall erratic), again reducing the effectiveness of RE. If those can be treated somehow, then convergence order is restored and RE performance improved. This is one more reason for applying the smoothing procedure described in Sec. 2.4.2.

## 3. Calibration

The main goal of the present work is to fit the GARCH diffusion model to a market of options. Some people choose to fit to option prices and others to the implied volatilities (IVs). We are strong proponents of the second approach. IV's are a natural way to regularize a set of option prices -- which can range from $0.05 to hundreds of dollars. IV's are the same order of magnitude across all the options. For SPX and other broad-based indices, using IV's will also weight higher the influence of deep out-of-the money puts. Given that such options are a difficult regime for models (especially diffusions) to fit well, we like this property as well. It stresses an area where models have difficulty.

Specifically, we try to fit the model to the option data by defining the following objective function to be minimized:

$$RMSE_{IV} = \sqrt{\frac{1}{N} \sum_{i=1}^{i=N}(IV_i^{model} - IV_i^{market})^2} = f(v_0, \bar{v}, \kappa, \xi, \rho, NS, NV, NT), \quad (31)$$

where $N$ is the number of options we wish to include in the calibration[14].

We calibrate to two SPX option chains, denoted Chain A and Chain B. Chain A used 246 SPX option quotes from Mar 31, 2017, filtered from quotes and IV's calculated by the CBOE. The data and notes for that are found in Appendix C.

For the optimization we use tools available in popular software[15]. We test two local optimizers, Excel's Solver tool which is based on the Generalized Reduced Gradient (GRG) method and Mathematica's FindMinimum function which is based on an interior point method. We also use Mathematica's NMinimize function, based on the global optimization Differential Evolution algorithm. All routines accept constraints which we impose on the model parameters in (31) in a way so that they encompass all plausible values.

To work with Excel and Mathematica[16], we build a dll exporting a function that returns the $RMSE_{IV}$ taking just the PDE engine's configuration as inputs. The function then reads the option chain data from a file, prices the options and evaluates (31). This is readily parallelized at the chain level, distributing the $N$ options across all available CPU cores. We apply some basic load balancing since resolution (and thus calculation time - roughly proportional to $(NS \times NV \times NT)$) may vary, as we discuss next.

---

[13] This excludes options with very small market prices which, as described in Sec. 3, will usually be priced with higher resolution than the nominal one input for the calibration.
[14] Obviously, NS, NV and NT (as well as the rest of the PDE engine's configuration like choice of scheme, etc) are kept constant throughout a calibration (except when a negative value is detected as explained next).
[15] While a more customizable solution integrated with the PDE engine would likely be made to converge faster, we wish to keep things simple here and focus mostly on the PDE engine.
[16] Calling the function from Mathematica is trivial using the .NET/Link.

For options of different expirations to be priced with similar accuracy, we need to have the number of time-steps $NT$ increasing with the expiration $T$. At the same time, the initial period of the valuation (close to the discontinuous initial conditions) always requires a minimum $NT$ to be resolved adequately. We roughly satisfy these requirements by taking the nominal $NT$ input in (31) to be the number of time steps per year for options with $T > 1$, i.e., we set $NT_{option} = \lfloor NT \cdot \max(T, 1) \rfloor$. We also find it is important to ensure that some minimum spatial resolution is used for the far out-of-the-money options in the chain, since those are more likely to incur higher relative pricing errors. More specifically, whenever the market value of an option is less than 0.5% of the asset spot $S_0$, $(NS \times NV)_{min}$ is set to $(120 \times 70)$ which is then gradually increased to $(400 \times 100)$ for market values of 0.01% of $S_0$ or lower. We've found these empirical choices lead to better efficiency in terms of obtaining more accurate fitted parameters faster.[17]

As was explained in Sec. 2.2, the present discretization allows for negative option values by design. In general, those occur when the resolution is too coarse (and thus the accuracy too low) and the correlation coefficient $\rho$ strongly negative. In practice we found such occurrences relatively rare under reasonable resolutions. With a local optimizer and when a previous result is used as the starting point, it is not unusual to complete a calibration involving tens of thousands of individual evaluations without a single negative value occurring. Such occurrences become even less frequent if one applies the transformation $X = \ln(S)$, i.e. discretizing and solving equation (3) instead of equation (2).

When negative valuations do occur during a calibration, the implied volatility cannot be calculated. In such cases we simply repeat the failed option valuation using our most robust (but least efficient) configuration, which involves switching to equation (3) and using the BDF3 scheme. We do so repeatedly, if required, using gradually increasing resolution until a positive value is returned. This "brute-force" approach can occasionally slow down a calibration, mostly when a global optimizer is used. On the other hand, it "automatically" ensures that the option is priced accurately, which wouldn't be the case if we used restrictions on the grid steps and/or added some sort of artificial diffusion aiming for an M-matrix (as was discussed in Sec. 2.2). Finally, since a valuation may just happen to be positive at $(S_0, v_0)$ (i.e., $V_0 > 0$), but still go significantly negative in the vicinity (and thus be inaccurate overall), the naïve check of $V_0 > 0$ is not sufficient. Instead, we check for negative values at all grid points within 10% of the strike in the $S$-direction and 50% of $v_0$ in the $v$-direction and discard any positive valuation $V_0$ if a negative value of magnitude more than 1% of $V_0$ is detected.

In general, if the model is to produce a decent fit to the market IV's (and thus prices), then as the optimizer homes in on the optimum parameter set, the chances of a model price being that close to zero and thus susceptible to this problem are very low.

## 3.1 Two approaches for the objective function evaluation

Given our KBE PDE solver, the first and most obvious approach to evaluating (31) is indeed to price each of the $N$ options separately, i.e., solve $N$ PDE's for each objective function evaluation. Each PDE is solved on a different grid based on $K, S_0$ and $T$, as described in Sec. 2.4.1. This is also the most general strategy since it can be used if we want to include options other than vanillas in the calibration exercise. The PDE engine could easily handle American or barrier options for example, at no extra cost. We shall refer to this as Approach I hereafter.

Our purpose here though is to calibrate to vanilla options, in which case we can make use of the scaling (MAP) property introduced in Sec. 1.1. This means that only one PDE solution is sufficient to provide all option prices (for all different strikes) in an expiration bucket $T_j$, $j = 1, \dots, N_E$, where $N_E$ is the number of different expirations included in the calibration. We choose to price one put option per $T_j$ and in particular the one that is furthest out-of-the-money (with the lowest strike price $K$). This put necessitates an $S$-grid with the highest required $S_{max}/K$ ratio (see Sec. 2.4.1.1) for our data[18]. The prices of the rest of the puts are then readily found via the scaling relation and interpolation on the $S$-grid. The call prices are obtained from the put price at the corresponding scaled $s_0$ via put-call-parity. Overall, this strategy (henceforth referred to as Approach II) requires $N_E$ PDE solutions for one evaluation of (31) and the actual $N$ option prices are extracted from those.

---

[17] It also ensures there are enough grid points where necessary for smooth higher order interpolation of the calculated option price at $(S_0, v_0)$, and helps to avoid negative values by enforcing some minimum accuracy.
[18] We include puts that are further out-of-the-money than the calls.

Naively thinking, one may expect this to result in $N/N_E$ times faster computation compared to Approach I (which would represent a 30-fold increase in the case of a chain with 240 options and 8 distinct expirations for example). In the previous section we described why for some options we want to use higher spatial resolution ($NS \times NV$). Each $T_j$ bucket may include one or more such options, seen as weak links in terms of computational efficiency. With Approach I this means that about 80% of the $N$ PDE's can be solved very fast on coarse grids[19], while the rest (a few options in each $T_j$ bucket) will be priced on a much finer grid and take longer. Allocating 240 PDE solvers across different CPU cores amounts to reasonable medium-grain parallelism and allows for decent load-balancing. With Approach II on the other hand these advantages are lost: if we want all the extracted option prices to be of equivalent accuracy to when calculated individually (as in Approach I), we need to account for the weakest link in each case. If each $T_j$ includes at least one option that requires a fine(r) grid, it follows that all $N_E$ PDE solvers need to use some overridden (high) resolution (as per previous section). This also means that now we may well have more CPU cores available than parallel tasks, say $N_E = 8$ and $N_{cores} = 10$, leaving some of the processing power unutilized. It may also lead to bad load balancing if for example one solver uses higher resolution that the rest. For this reason, we lower the maximum enforced $(NS \times NV)_{min}$ from $(400 \times 100)$ for Approach I to $(200 \times 80)$ for Approach II. The lesser accuracy that this implies for the few deep out-of-the-money options is offset by the fact that now all option prices are extracted from fine grids (as opposed to only a few with Approach I). As we will see next, this leads to calibrated model parameter accuracy as high as under Approach I.

For the ADI schemes there is an alternative strategy and that is to parallelize at the PDE solver level. This is because grid lines can be updated simultaneously during both the explicit and implicit steps. Our brief tests with 8 or 10 cores/threads show that even with basic OpenMP instructions, a parallel efficiency of 80% is readily achievable this way, resulting in similar calibration times to our main, chain-level parallelization approach.

## 4. Numerical experiments and results

We now analyze the performance of the PDE engine and calibrator and present results using our two sample SPX option datasets, 246-option Chain A representing the 2017 low-volatility market, and 68-option Chain B from the higher volatility environment of 2010. Chain A data are given in Appendix C. The timings were taken on a 10-core Intel i9-7900X PC; the code was written in C++ and compiled in VS2013. We perform most of our tests using Approach I of the previous section (i.e. pricing each option separately). We do so since it is obviously preferable to assess the behavior/performance of the PDE engine over a sample of 246 or 68, rather than 8 or 7 individual option evaluations.

### 4.1. General findings

The ADI methods, as expected, prove to be more efficient than the BDF3 scheme and can be depended upon for successful and fast calibrations. At an individual option pricing level though we found they are less robust as they are not immune to spurious oscillations, mostly in the delta and gamma of the solution. The most likely offender is the HV1 scheme (which also proves to be the most efficient overall). We found this problem much rarer with the MCS scheme for the vanilla payoffs we're dealing with here[20]. The fully implicit BDF3 scheme demonstrates superior damping properties; we have been unable to reproduce a single case of this problem in extensive testing. In practice, with all reasonable $(NS, NV, NT)$ the ADI schemes are oscillation-free as well. Even when trying to force the issue (using some low $NT/NS$), our tests show that any such mild oscillations present in the individual solutions do not usually prevent optimizer convergence (but they do slow it down). We note that this is a more general feature: the less accurate the PDE solution is, the more steps are usually required for the optimizer to converge. Using too low NT for example, will cause the optimizer to "hunt" more; conversely, application of (spatial) Richardson extrapolation typically means convergence in fewer steps than when simple (non-extrapolated) solutions are used.

---

[19] We will see in the next section that a resolution of $(NS \times NV) = (60 \times 30)$ is more than enough to obtain accurate IV's for the bulk of the options that are not deep out-of-the-money.
[20] Wyns [9] shows that such problems are more common when the MCS scheme is used to price cash-or-nothing options, where the discontinuity in the initial conditions is much more severe.

The other relevant problem is that of the occasional negative values. The use of equation (3) results in non-negative solutions under moderate spatial grid resolutions in cases where this is impossible with equation (2) and any reasonable resolution. Moreover, we find that in such cases the BDF3 scheme does a better job than the ADI schemes. As an example, we mention in passing a particularly difficult case that arose during a calibration with the global optimizer. In that case $NT = 20$ was enough for the BDF3 scheme to produce a smooth, non-negative price profile near the strike, whereas the ADI schemes needed $NT > 20000$ to achieve the same result (with the same spatial discretization).

In terms of the optimizers we tested, we found Excel's Solver tool to be the best choice. The main reason is that it benefits from a good initial guess, whereas Mathematica's FindMinimum does not. The Solver will typically converge 2.5 to 5 times faster when the starting vector is not too far from the optimal. Despite both being local optimizers, we are confident that they can be used to find the true (global) minimum. Extensive testing using many different starting points shows that both converge to the same vector. Using Mathematica's NMinimize global optimization routine further confirmed the solution in every case we tested. NMinimize also served as a torture test for the PDE engine, exploring all corners of the parameter space. All schemes never failed to produce a valid price, though of course "difficult" parameter sets (and insufficient resolution) generally trigger the repricing mechanism. Even so, the total optimization time is not significantly affected in practice. The allowed parameter ranges we used for the tests were: $0.0025 \leq v_0 \leq 0.50$, $0.005 \leq \bar{v} \leq 0.25$, $1 \leq \kappa \leq 20$, $1 \leq \xi \leq 20$, $-0.95 \leq \rho \leq 0$, which cover most market scenarios.

## 4.2 PDE engine tests - Pricing

To test the convergence behavior of the time-marching schemes, we fix the spatial resolution to $(NS \times NV) = (60 \times 30)$ and calculate (time-converged) benchmark prices using the BDF3 scheme with $NT = 12800$. We also apply spatial Richardson extrapolation[21]. This way the spatial discretization error is low, but not negligible compared to the temporal error. Nonetheless, the two errors are found to be only weakly dependent, allowing the comparative performance of the schemes to be properly assessed. The prices are obtained via an objective function evaluation under Approach I, which means that every option in the chain is priced individually and under the resolution overriding / repricing rules described in Sec. 3. The pricing errors for various $NT$ are calculated as the differences from the benchmark prices and the RMSE is used as an indicator of the overall performance for each scheme.

Figure 1 shows the results for the HV1, HV2, MCS and BDF3 schemes, plus the HV1 scheme with Rannacher time-stepping (hereafter referred to as HV1D). The points on the left correspond to practical $NT$ (and CPU times) while those on the right are included to better illustrate the asymptotic behavior[22]. We plot the relative (as opposed to absolute) pricing errors, since those are more closely related to the errors in the implied volatilities and consequently the calibrated model parameters.

The HV1, HV2 and MCS schemes display a linear relationship between RMSE and CPU time on the logarithmic scale. This reflects (and confirms) their theoretical second-order convergence and the fact that the execution time in their case is proportional to $NT$. The Implicit Euler damping step (which requires an expensive factorization of the full system matrix) introduces an upfront cost that lowers the efficiency of the HV1D scheme. The irregular first two points from the left of the HV2 curve for Chain A are an example of the repricing mechanism in action: the scheme's accuracy is too low here, causing some options in the chain to fail the "negative values test" (which are then revalued with a different configuration). Finally, for the BDF3 scheme we have an upfront cost that just like with HV1D is due to the initial matrix factorizations, resulting in significantly reduced efficiency for practical $NT$. At large $NT$ this effect is diluted, and the scheme is seen to confirm its theoretical third-order convergence.

---

[21] The (composite) RE solution combines solutions on the $(NS \times NV)$ and $(NS/2 \times NV/2)$ grids.
[22] The rightmost points of the error curves in Figure 1 correspond to $NT = 1600$ for the ADI schemes and $NT = 400$ for the BDF3.

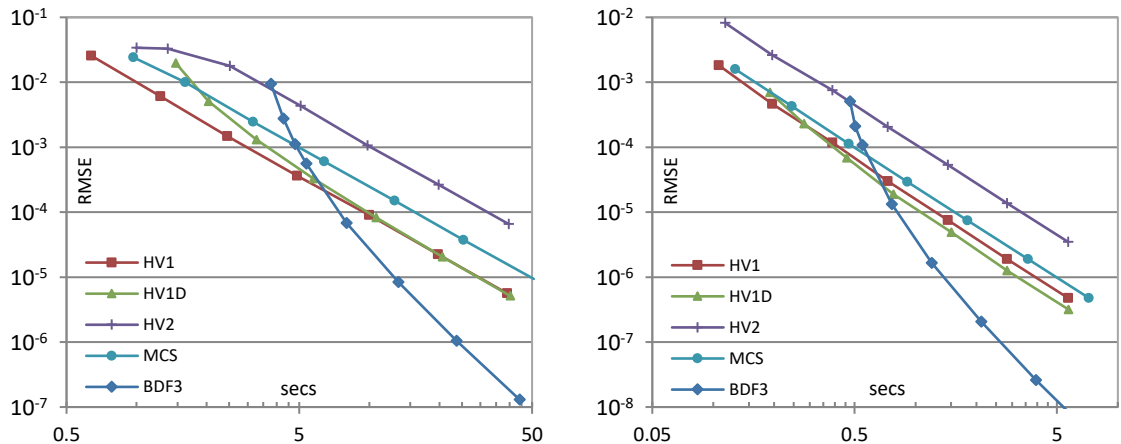

**Figure 1**. Computational efficiency of time-marching schemes. RMS relative temporal error vs CPU time for the pricing of Chain A (246 options, left) and Chain B (68 options, right). Model parameters from Table 1.

Overall, the HV1, HV1D and MCS schemes offer comparable performance. Based on the present and many more similar tests, the HV1 scheme may well be the best choice. Under typical usage, any spurious oscillations do not seem to significantly affect its calibration performance. Adding damping comes at a mild cost overall and the HV1D scheme in fact produces lower (relative) errors for Chain B[23]. The MCS scheme is still the most robust of the considered ADI schemes[24] and usually about as accurate as HV1/HV1D, but it is slightly more computationally expensive per time-step. The HV2 scheme is the worst-performing and we will not consider it here further. Finally, the BDF3 scheme perhaps fares better than expected, beginning to outperform the ADI schemes for relative errors below around 0.02% in the case of Chain A. It is interesting to note that all temporal discretization errors are about an order of magnitude smaller for Chain B, making the BDF3 scheme less competitive in this case. Overall, we (unsurprisingly) find it cannot match the best ADI schemes' efficiency for practical accuracy goals.

## 4.3 PDE engine tests – Calibration (Approach I)

We now test the performance of the PDE engine in terms of the end-result, the calibrated model's parameters. Benchmark values are given in Table 1, calculated using the BDF3 scheme and a resolution of $(NS \times NV \times NT) = (800 \times 200 \times 50)$ using spatial RE.

**Table 1.** Benchmark calibrated model parameters

|  | $v_0$ | $\bar{v}$ | $\kappa$ | $\xi$ | $\rho$ |
|---|---|---|---|---|---|
| Option chain A (Mar 31, 2017) | 0.010935 | 0.039139 | 5.3905 | 6.8997 | -0.74579 |
| Option chain B (Feb 1, 2010) | 0.044816 | 0.088529 | 3.6695 | 5.0333 | -0.79206 |

Figure 1 suggests that the HV1 scheme is more efficient than the MCS scheme for Chain A, but the two schemes perform very similar for Chain B. This translates to the calibrated model parameters. Table 2 shows that the parameters obtained for Chain A with the HV1 scheme are more converged for any $NT$ than those obtained with the MCS scheme and the CPU time required for the calibration is lower as well. We note that in this case the BDF3 scheme with $NT = 25$ gives about the same accuracy as the

---

[23] The absolute errors (not shown) are slightly lower as well compared to the non-damped HV1, for both chains A and B. This may indeed suggest the presence of mild oscillations with the HV1 scheme (resulting in lowered accuracy) that are successfully suppressed by the added damping of the HV1D scheme.

[24] We've found cases where the Euler damping (Rannacher time-stepping) procedure of the HV1D scheme reduces but doesn't eliminate spurious oscillations, whereas for the same cases the MCS scheme is oscillation-free (even without damping). In contrast, we have not seen any cases where the reverse is true.

HV1 scheme with $NT = 200$ and requires the same time as well. Table 3 confirms the two ADI schemes produce similar accuracy for the Chain B parameters. The latter are also more converged for all $NT$ compared to the Chain A parameters in Table 2, reflecting the lower overall errors on the right plot of Figure 1. As in Sec. 4.2, we used a fixed spatial resolution of $(NS \times NV) = (60 \times 30)$ with (spatial) RE and obtained time-converged calibrated model parameters using the BDF3 scheme with $NT = 400$.

**Table 2.** Model parameter convergence for calibrations to Chain A (246 options) as a function of temporal resolution for the HV1, MCS and BDF3 schemes. The exact (time-converged) parameter set is $(v_0, \bar{v}, \kappa, \xi, \rho) = (0.010937, 0.039131, 5.3914, 6.9010, -0.74567)$.

|            | HV1      |          |          |          | MCS      |          |          |          | BDF3     |          |
|------------|----------|----------|----------|----------|----------|----------|----------|----------|----------|----------|
| **NT**     | 25       | 50       | 100      | 200      | 25       | 50       | 100      | 200      | 25       | 50       |
| $v_0$      | 0.010872 | 0.010920 | 0.010933 | 0.010936 | 0.010864 | 0.010912 | 0.010931 | 0.010936 | 0.010937 | 0.010937 |
| $\bar{v}$  | 0.039322 | 0.039178 | 0.039142 | 0.039134 | 0.039302 | 0.039207 | 0.039150 | 0.039136 | 0.039128 | 0.039131 |
| $\kappa$   | 5.3518   | 5.3821   | 5.3891   | 5.3908   | 5.3635   | 5.3764   | 5.3876   | 5.3904   | 5.3920   | 5.3915   |
| $\xi$      | 6.9480   | 6.9130   | 6.9039   | 6.9017   | 6.9497   | 6.9223   | 6.9061   | 6.9023   | 6.8996   | 6.9009   |
| $\rho$     | -0.73655 | -0.74338 | -0.74511 | -0.74554 | -0.73753 | -0.74197 | -0.74476 | -0.74545 | -0.74586 | -0.74570 |
| CPU (mm:ss)| 01:52    | 03:20    | 06:28    | 12:43    | 03:08    | 04:22    | 08:20    | 16:02    | 12:17    | 18:20    |

**Table 3.** Model parameter convergence for calibrations to Chain B (68 options) as a function of temporal resolution for the HV1, MCS and BDF3 schemes. The exact (time-converged) parameter set is $(v_0, \bar{v}, \kappa, \xi, \rho) = (0.044816, 0.088478, 3.6726, 5.0329, -0.79233)$.

|            | HV1      |          |          |          | MCS      |          |          |          | BDF3     |          |
|------------|----------|----------|----------|----------|----------|----------|----------|----------|----------|----------|
| **NT**     | 25       | 50       | 100      | 200      | 25       | 50       | 100      | 200      | 25       | 50       |
| $v_0$      | 0.044800 | 0.044812 | 0.044815 | 0.044815 | 0.044805 | 0.044813 | 0.044815 | 0.044815 | 0.044814 | 0.044815 |
| $\bar{v}$  | 0.088482 | 0.088480 | 0.088479 | 0.088478 | 0.088508 | 0.088487 | 0.088481 | 0.088479 | 0.088480 | 0.088479 |
| $\kappa$   | 3.6781   | 3.6738   | 3.6729   | 3.6727   | 3.6745   | 3.6729   | 3.6726   | 3.6726   | 3.6731   | 3.6727   |
| $\xi$      | 5.0403   | 5.0346   | 5.0333   | 5.0330   | 5.0394   | 5.0344   | 5.0332   | 5.0329   | 5.0336   | 5.0330   |
| $\rho$     | -0.79120 | -0.79205 | -0.79226 | -0.79232 | -0.79121 | -0.79204 | -0.79226 | -0.79231 | -0.79227 | -0.79233 |
| CPU (mm:ss)| 00:20    | 00:40    | 01:16    | 02:27    | 00:26    | 00:52    | 01:39    | 03:10    | 01:56    | 02:42    |

A careful inspection of the sequence of parameters obtained with the two ADI schemes with increasing (doubling) $NT$, reveals very close to second-order convergence for $NT = 50$ and above, indicating that the theoretical order of the time-discretization translates to the "functional" of the computation, i.e., the model parameter vector. This suggests the possible use of (temporal) Richardson extrapolation on the fitted parameters. The effectiveness of such an approach is demonstrated in Table 4, where the parameter vectors have been obtained as the composite (extrapolated) result of two successive calibrations using slightly different $NT$. The vector from the first calibration can be used as the starting point for the second, reducing the time of the latter[25].

Comparing for example the $NT = 100$ calibrations in Table 2 with the (composite) $NT = (40,60)$ calibrations in Table 4 for Chain A, one can see that the parameters obtained with the latter are significantly more converged, while the CPU times are lower as well. As always with RE, care should be taken not to use too low a resolution (in this case $NT$)[26]. To guarantee good RE performance we recommend using $NT_{coarse} \geq 30$ and $NT_{fine}/NT_{coarse} \geq 1.2$.

The effect of spatial RE[27] on the calibrated parameters is shown in Tables 5 & 6. Here we use the BDF3 scheme with $NT = 50$ so that the temporal discretization error is negligible. The benchmark parameters are from Table 1. Despite resolution being low, the parameters obtained with RE are not far from the benchmark values. For Chain A they are practically converged to 4 digits (the error is at most one point off in the fourth digit), while for Chain B they are somewhat less converged. In both cases the parameters obtained without the use of RE are significantly less accurate.

---

[25] Unless otherwise noted, all calibration CPU times listed in the present work are obtained with Excel's Solver starting from the average values vector $(v_0, \bar{v}, \kappa, \xi, \rho) = (0.05, 0.05, 5, 5, -0.7)$.
[26] Also, the ADI schemes may not always sufficiently damp oscillations when $NT/NS$ is too low, possibly leading to erratic convergence and thus RE results. This is mainly an issue for the HV1 scheme.
[27] Applied on the PDE solution (option prices) as described in Sec. 2.4.3.

Table 4. Model parameter convergence using the ADI schemes and temporal Richardson extrapolation on the fitted parameters. $(NS \times NV) = (60 \times 30)$ with spatial Richardson extrapolation. The exact (time-converged) parameter set is $(v_0, \bar{v}, \kappa, \xi, \rho) = (0.010937, 0.039131, 5.3914, 6.9010, -0.74567)$ for Chain A and $(v_0, \bar{v}, \kappa, \xi, \rho) = (0.044816, 0.088478, 3.6726, 5.0329, -0.79233)$ for Chain B.

|  | Chain A (246 options) | | | | Chain B (68 options) | | | |
| --- | --- | --- | --- | --- | --- | --- | --- | --- |
|  | HV1 | | MCS | | HV1 | | MCS | |
| NT | 40,60 | 80,100 | 40,60 | 80,100 | 30,36 | 60,80 | 30,36 | 60,80 |
| $v_0$ | 0.010937 | 0.010937 | 0.010938 | 0.010937 | 0.044816 | 0.044816 | 0.044816 | 0.044816 |
| $\bar{v}$ | 0.039130 | 0.039131 | 0.039131 | 0.039131 | 0.088479 | 0.088478 | 0.088481 | 0.088479 |
| $\kappa$ | 5.3918 | 5.3914 | 5.3910 | 5.3914 | 3.6724 | 3.6726 | 3.6723 | 3.6725 |
| $\xi$ | 6.9008 | 6.9009 | 6.9005 | 6.9009 | 5.0327 | 5.0328 | 5.0328 | 5.0328 |
| $\rho$ | -0.74571 | -0.74569 | -0.74567 | -0.74568 | -0.79233 | -0.79233 | -0.79231 | -0.79233 |
| CPU (mm:ss) | 05:07 | 09:07 | 07:07 | 11:20 | 00:52 | 01:08 | 01:00 | 01:20 |

Table 5. Effect of spatial Richardson extrapolation on model parameter convergence for Chain A.

| Spatial resolution | $v_0$ | $\bar{v}$ | $\kappa$ | $\xi$ | $\rho$ |
| --- | --- | --- | --- | --- | --- |
| $(NS \times NV) = 60 \times 30$ | 0.010914 | 0.039232 | 5.2805 | 6.8705 | -0.74333 |
| $(NS \times NV) = 60 \times 30$ w RE | 0.010937 | 0.039131 | 5.3915 | 6.9009 | -0.74570 |
| Benchmark | 0.010935 | 0.039139 | 5.3905 | 6.8997 | -0.74579 |

Table 6. Effect of spatial Richardson extrapolation on model parameter convergence for Chain B.

| Spatial resolution | $v_0$ | $\bar{v}$ | $\kappa$ | $\xi$ | $\rho$ |
| --- | --- | --- | --- | --- | --- |
| $(NS \times NV) = 60 \times 30$ | 0.044811 | 0.088663 | 3.6064 | 5.0181 | -0.79212 |
| $(NS \times NV) = 60 \times 30$ w RE | 0.044815 | 0.088479 | 3.6727 | 5.0330 | -0.79233 |
| Benchmark | 0.044816 | 0.088529 | 3.6695 | 5.0333 | -0.79206 |

## 4.4 Faster calibration using the MAP property (Approach II)

So far, we have presented calibration tests where every option in the chain is priced separately. These tests demonstrated the efficiency of the PDE engine; calibrations using this Approach are already quite fast. We now test Approach II for the objective function evaluation, i.e., making use of the MAP property. Following the winning combination of Table 4, we choose the HV1 scheme, a resolution of $(NS \times NV) = (60 \times 30)$[28] with spatial RE, as well as temporal RE on the fitted parameters (combining the results of two successive calibrations) with NT = (30,36).

As expected, Table 7 confirms that the speed-up compared to Approach I is significant, especially for the largest Chain A[29]. The parameter accuracy is at least as good. It is obvious that this approach works well in practice and effectively decouples the calibration time from the total number of options included. For either of our datasets a CPU time of less than a minute is needed to achieve a maximum relative (numerical) error of 0.05% for the obtained parameters. We stress that a judicious *S*-grid construction (low to moderate non-uniformity) is key for keeping the solution error profile low across the moneyness spectrum and make this approach work. As we already mentioned in Sec. 4.1, lower overall pricing accuracy leads not only to less accurate parameters but also (perhaps more importantly) to slower convergence of the optimizer.

---

[28] While this nominal resolution is used by most of the PDE solvers within Approach I, in the case of Approach II, all (7 or 8) solvers really use higher resolution, as explained in Sec. 3.1.
[29] We note that the code was developed on an older CPU (4-core Intel i7-920, 2009) and not the new 10-core i9-7900X CPU used for the timings reported here. The MAP performance gains on the development CPU are almost double (6×) those of the newer CPU, and the maximum calibration time still around 2 mins.

**Table 7.** Comparison of pricing Approach I (one PDE solution per option) and Approach II (one PDE solution per expiration). The relative errors (compared to the benchmark parameters of Table 1) are shown in parentheses.

|  | Chain A (246 options) | | Chain B (68 options) | |
| --- | --- | --- | --- | --- |
|  | Approach I | Approach II | Approach I | Approach II |
| $v_0$ | 0.010937 (0.01%) | 0.010940 (0.05%) | 0.044816 (0.00%) | 0.044818 (0.01%) |
| $\bar{v}$ | 0.039128 (0.03%) | 0.039135 (0.01%) | 0.088479 (0.06%) | 0.088505 (0.03%) |
| $\kappa$ | 5.3923 (0.03%) | 5.3906 (0.00%) | 3.6724 (0.08%) | 3.6712 (0.05%) |
| $\xi$ | 6.9010 (0.02%) | 6.8988 (0.01%) | 5.0327 (0.01%) | 5.0323 (0.02%) |
| $\rho$ | -0.74569 (0.01%) | -0.74581 (0.00%) | -0.79233 (0.03%) | -0.79209 (0.00%) |
| CPU (mm:ss) | 00:03:00 | 00:00:55 | 00:00:52 | 00:00:40 |

## 4.5 More calibration results and comparison to the Heston model

We now present detailed calibration results demonstrating the ability of the GARCH diffusion model to fit the option market and compare it to the popular Heston model. The Heston calibrations were performed using the present PDE engine[30] and the resulting parameter vectors were then confirmed with independent calibrations using pricing via well-known Fourier integral representations. Figures 2 and 3 illustrate the market fit for each option expiration bucket in chains A and B respectively. As a first remark we can say that the model is indeed able to capture a smile behavior in the short end and achieve an overall decent fit. (By 'smile behaviour', we mean that the IV curve has an evident minimum). This can be seen in the first two plots in Figure 2, but not in Figure 3 (the Heston model can be seen to capture the smile in both cases). The reason is that the strike point *K\** (where the GARCH diffusion model's IV curve turns up) for the first two expirations in Chain B lies further to the right of the last market point included in the plot (at around *K* = 1225).

Overall, both sample calibrations seem to indicate that the Heston model is more 'flexible', managing to fit the data better overall. The GARCH diffusion model fits on the other hand look more 'rigid'. This is somewhat surprising and in contrast with the findings of Christoffersen et al. [1]. While our two-chain data set is tiny compared to theirs, we suspect the contrast in findings is due to our much wider (smile) moneyness coverage – as no downside puts are used in [1].

The apparent Heston model victory here comes with known problems. The very small obtained Feller ratios, $R \equiv 2\kappa\bar{v}/\xi^2$, (0.12 and 0.29) are well below one. Note that under S. Heston's 1993 model [14], R is the same under both P (physical) and Q (risk-neutral) model evolutions. In our experience, the Heston P-model estimates (from time series, using maximum likelihood) will typically have $R > 1$. There are some caveats to complaining about *R*: for either model, P-model parameter estimates are not trivially obtained because the latent volatility must either be proxied or jointly estimated. In addition, P-model time series estimates are typically quite sensitive to the inclusion or not of crash days like Oct. 19, 1987. [31]

---

[30] To trivially adapt the PDE engine to the Heston pricing PDE: adjust the *v*-diffusion and mixed derivative coefficients $d_v$ and $m_{Sv}$ accordingly in (17) and (20). The only other change required is the choice of $v_{max}$.

[31] We believe the 'hitting range', $0 < 2\kappa\bar{v}/\xi^2 \leq 1$, should be admitted in calibrations. However, once you find the optimal Q-estimate ratio in the hitting range, you are forced to ponder the implications. Indeed, the volatility distribution then develops an integrable divergence at zero, v = 0 becomes the most probable value, and repeated volatility hits on zero become possible. If the P-evolution model has $1 < 2\kappa\bar{v}/\xi^2$, arbitrage opportunities develop – at least in the idealized continuous-time world in which the models are constructed. Yet, finding a smile-calibrated Feller ratio in the hitting range is well-known to be a common occurrence [15].

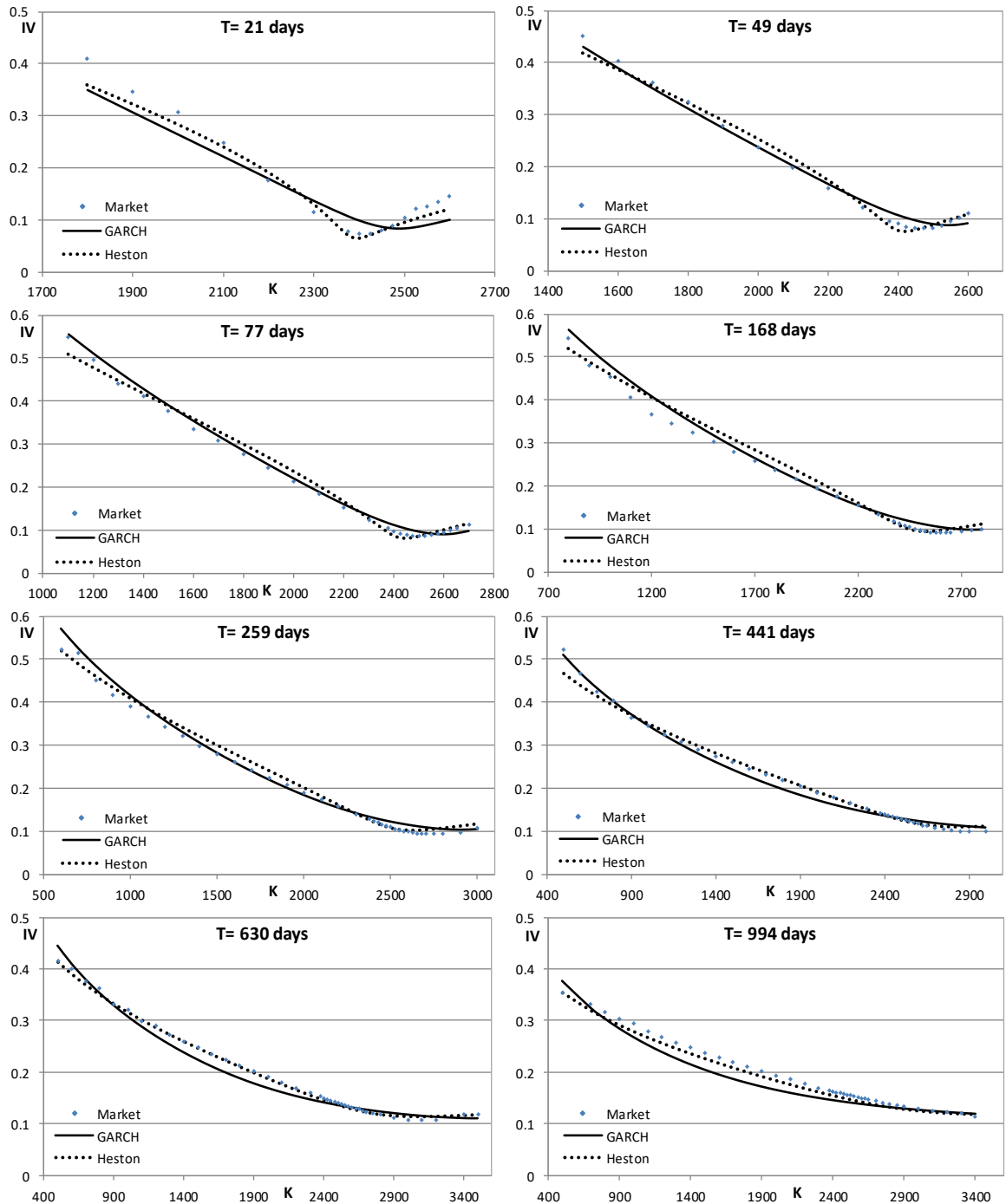

**Figure 2**. GARCH diffusion and Heston model implied volatilities by expiry for Chain A. The GARCH diffusion model parameters are given in Table 1. The parameters for the Heston model are $v_0 = 0.007316$, $\bar{v} = 0.03608$, $\kappa = 6.794$, $\xi = 2.044$ and $\rho = -0.7184$. GARCH diffusion $\text{RMSE}_{IV} = 1.68\%$, Heston $\text{RMSE}_{IV} = 1.28\%$. Heston Feller ratio = 0.12.

To purposely exaggerate the Feller ratio issue, we also fitted only the first two shortest expirations in Chain A. Despite seemingly achieving a decent fit (Figure 4), the Heston model fitted $v_0$ is practically zero and the Feller ratio is 0.05. The GARCH diffusion model on the other hand achieves a closer fit with mostly reasonable parameters and a volatility process that doesn't hit zero. Nevertheless, our calibrated volatility-of-volatility $\xi$-values for the GARCH diffusion are also likely 'too high', relative to typical P-estimates. In the case of Chain B (Figure 5) the Heston model does a slightly better (if oscillatory) job, following the smile correctly (with the model $K^*$ very close to the market $K^*$ at around $K = 1140$), while the GARCH diffusion model $K^*$ is about 1180.

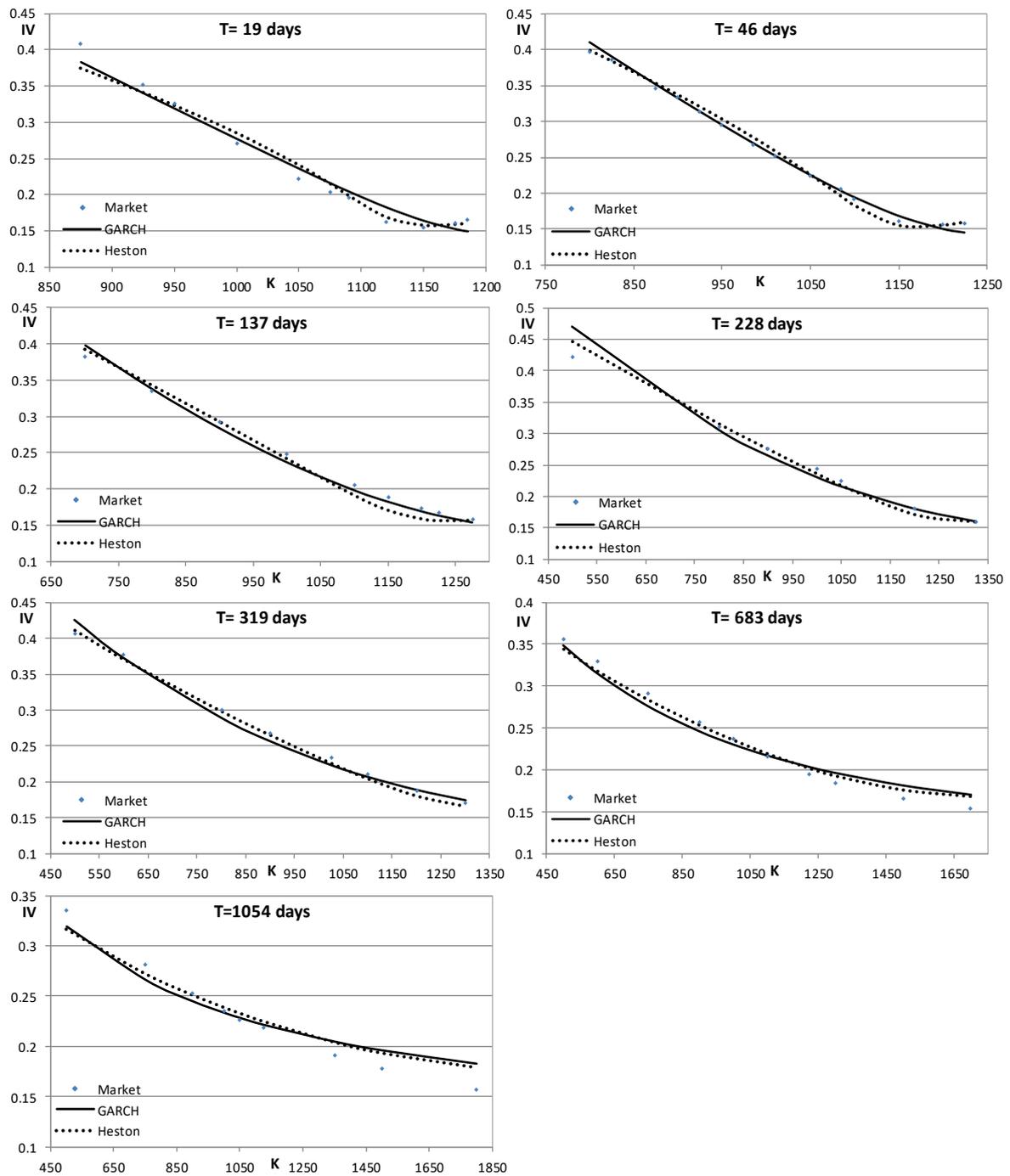

**Figure 3**. GARCH diffusion and Heston model implied volatilities by expiry for Chain B. The GARCH diffusion model parameters are given in Table 1. The parameters for the Heston model are $v_0 = 0.04576$, $\bar{v} = 0.06862$, $\kappa = 4.905$, $\xi = 1.525$ and $\rho = -0.7131$. GARCH diffusion $\text{RMSE}_{IV}$ = 1.19%, Heston $\text{RMSE}_{IV}$ = 1.01%. Heston Feller ratio = 0.29.

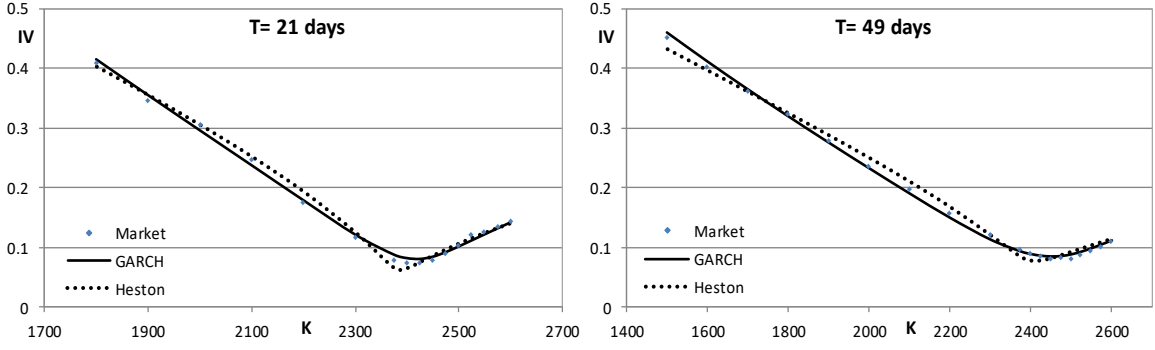

**Figure 4**. GARCH diffusion and Heston model implied volatilities when fitted only to the first two expiries in Chain A. The GARCH model parameters are $v_0 = 0.008046$, $\bar{v} = 0.02981$, $\kappa = 10.93$, $\xi = 15.06$ and $\rho = -0.5669$. The Heston model parameters are $v_0 = 10^{-6}$, $\bar{v} = 0.02$, $\kappa = 40.64$, $\xi = 5.65$ and $\rho = -0.623$. GARCH diffusion $\text{RMSE}_{IV} = 0.61\%$, Heston $\text{RMSE}_{IV} = 0.86\%$. Heston Feller ratio = 0.05.

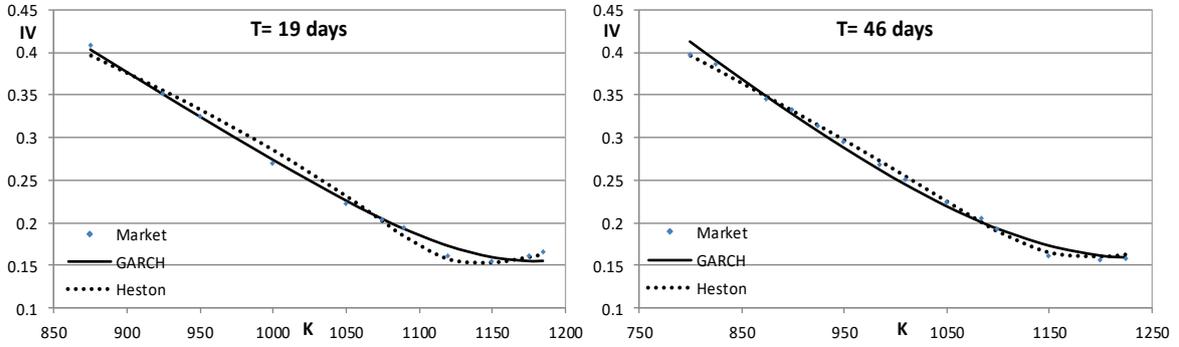

**Figure 5**. GARCH diffusion and Heston model implied volatilities when fitted only to the first two expiries in Chain B. The GARCH model parameters are $v_0 = 0.03836$, $\bar{v} = 0.07067$, $\kappa = 11.96$, $\xi = 7.685$ and $\rho = -0.6871$. The Heston model parameters are $v_0 = 0.03641$, $\bar{v} = 0.06011$, $\kappa = 21.3$, $\xi = 2.604$ and $\rho = -0.6637$. GARCH diffusion $\text{RMSE}_{IV} = 0.64\%$, Heston $\text{RMSE}_{IV} = 0.59\%$. Heston Feller ratio = 0.38.

To summarize, at the outset, we knew both models have their nice properties and their issues. In particular, neither model handles extreme moves – in either the asset price or its underlying volatility – either naturally or well. However, although our dataset is very small, we were still surprised to see the Heston model achieve the better smile fits.

## 4.6 Some easy PDE extensions

Both models we've calibrated so far are subcases of the more general power-law SV model

$$dS_t = (r_T - q_T)S_t dt + \sqrt{v_t} S_t dW_t^S,$$
$$dv_t = \kappa(\bar{v} - v_t)dt + \xi v_t^p dW_t^v. \quad (32)$$

The popular affine Heston model ($p = 0.5$) has some well-known limitations, such as: (a) inability to capture steep short-term volatility smiles, (b) instability of the fitted parameters over recalibrations and (c) incompatibility of fitted parameters with those estimated from the P-world. The GARCH diffusion model ($p = 1$) has received less attention in the literature, but our tests here indicate that it too suffers from (a). Regarding (b) and (c) more tests would be needed; it may well be that GARCH produces more stable parameters than Heston for example. In an effort to address these issues, researchers have introduced a variety of ideas. Again, naming some: (i) moving away from affine models, so using the more general *p*-model above (or other two-factor diffusion variations), (ii) randomizing the (latent) spot variance $v_0$, (iii) adding jumps, (iv) adding more stochastic factors, (v) using fractional Brownian motion as the volatility driver. Among these extensions, (i) and (ii) are particularly simple to add to the present framework, so we will briefly explore them here.

Our PDE engine can easily solve for either the GARCH diffusion, Heston, or any model "in between", i.e., with $0.5 \leq p \leq 1$. It is in fact straightforward to let $p$ float and have the optimizer decide its optimal value. As Figures 2 and 3 show, a value of $p = 0.5$ corresponds to more 'flexible', whereas $p = 1$ to more 'rigid' fits. Intermediate values of $p$ predictably yield fits that look like a mix between the two. A low value of $p$ reduces the $RMSE_{IV}$ by locally enabling enough curvature to emulate the short-term market smiles and on the other hand it increases it as the (overly) flexible behavior persists for the middle expirations (Heston case). Overall though, at least for our datasets, lower $RMSE_{IV}$ is achievable with lower values of $p$. Our tests found an optimum $p$ of 0.62 for Chain A ($RMSE_{IV}$ = 1.23%), while for Chain B we found $p = 0.59$ ($RMSE_{IV}$ = 1%). As a result, the improvement in the fit was not significant over that of the Heston model (relative $RMSE_{IV}$ reduction of 1% - 4%).

This may not be telling the whole story though. Let's accept that both Heston and GARCH diffusion are essentially misspecified for the cause. Because of that, the optimizer is forced into unnatural parameters (e.g. unrealistically low $v_0$ and/or high $\xi$) to adapt to the observed short-term smiles. If those two models are misspecified, so will be the general $p$-model. Our conclusion is that any benefit from choosing a particular value of $p$ for the general $p$-model, would be best assessed if some sort of short-term-smile-enabling extension was in place, leaving the diffusion model more freedom to fit the rest of the expirations. Normally this would involve adding jumps, but an easier way is available through randomization; see Mechkov [16] or Jacquier & Shi [17]. While we find the dynamic rationale of this approach perhaps not entirely clear, we also found it does succeed in "turning on" (up) the short-term smiles. Since again the changes needed in our code to add this feature were minimal, we gave it a try. Table 8 presents summary $RMSE_{IV}$ results for Chain A. As we noted above, varying $p$ when the model lacks the ability to match the short-term smile makes little difference. But when the smile is better accounted for (here via the randomization), the optimal $p$-model provides a more substantial improvement in the calibration fit over both the Heston and the GARCH diffusion model. This case is presented in more detail in Appendix D.

We finally add that other non-affine two-factor variations can also easily be accommodated. As an example, we calibrated the Inverse Gamma SV model introduced by Langrené et al. [15]. As can be seen in Table 8, it placed between the Heston and GARCH diffusion models in terms of overall quality of fit ($RMSE_{IV}$ for Chain B was 1.11%). We also note is that the fitted $v_0$ and $\bar{v}$ for this model were quite close to each other for both datasets (not shown).

**Table 8.** Calibration fit for Chain A under variations of pure diffusion models.

| Model | $RMSE_{IV}$ |
|---|---|
| Power law models | |
| Heston | 1.28% |
| GARCH diffusion | 1.68% |
| Optimal $p$-model | 1.23% |
| Heston - Randomized | 0.96% |
| GARCH diffusion - Randomized | 0.94% |
| Optimal $p$-model - Randomized | 0.79% |
| Inverse Gamma Vol model | 1.53% |

# 5. Conclusions

In this work we present a first (to our knowledge) full option calibration of the GARCH diffusion model using a PDE approach. The calibration is very fast and accurate (less than a minute on a modern PC) ameliorating the lack of a closed-form solution. This is accomplished with the use of an efficient yet "ordinary" second-order finite difference based PDE engine. While here we calibrate to European vanilla options, the same pricing engine can be used with only minor modifications for fast calibrations to other types of options that can easily be handled in the PDE setting, e.g. American options or barriers. Other similar models can also easily be accommodated as our brief experiments of Sec. 4.6 show.

In a small test with two SPX option chains, the smile fits with the GARCH diffusion model were inferior to the fits from the Heston '93 model. This differed from some prior literature such as [1]. The Heston fits come with very low values for the so-called Feller condition ratio, which leads to other issues. Nevertheless, we were surprised.

Our more general contribution is showing closed-form solutions need not be such a strong criterion for model selection. Similar PDE engines can potentially handle various related models that are being largely ignored in practice and therefore allow a more informed choice for a particular trading area.

For the future, it would be quite interesting to extend the solver to one that could handle bivariate jump-diffusions with similar high efficiency.

Finally, the second author (Lewis) would like to stress that the first author (Papadopoulos) has done all the heavy lifting here: developing and implementing all the C/C++ solvers and their Excel and Mathematica interfaces.

# Appendix A – Market prices of risk and a compatible real-world evolution

We have performed option chain calibrations after postulating that the risk-neutral (aka Q-measure) evolution has the GARCH diffusion form (1). Jumping immediately to a risk-neutral model is a common finance short-cut. More carefully, even given a target Q-model, one should begin with a compatible real-world (P-measure) evolution, and then move to the desired Q-measure evolution by a Girsanov transformation. In the presence of a deterministic stock dividend yield $q_t$, a P-measure evolution compatible with a Q-measure GARCH diffusion under this procedure has the form:

$$dS_t = (\alpha_t^P - q_t)S_t dt + \sigma_t S_t dW_{P,t}^S,$$
$$dv_t = \beta_t^P dt + \xi v_t dW_{P,t}^v.$$

Now $(W_{P,t}^S, W_{P,t}^v)$ are a pair of correlated P-Brownian motions with correlation $\rho$, both $(\rho, \xi)$ are identical under P or Q, and we used $\sigma_t \equiv \sqrt{v_t}$. Indeed, "no-arbitrage" requires that the Q-evolution must be related to the P-evolution by the Girsanov substitutions $dW_{P,t}^S = dW_{Q,t}^S - \lambda_t^e dt$ and $dW_{P,t}^v = dW_{Q,t}^v - \lambda_t^v dt$. Under this implied change-of-measure, the variance-covariance structure of the SDE is preserved but the drifts may change. Financially, $(\lambda_t^e, \lambda_t^v)$ represent market prices of (equity, volatility) risk. The $\lambda$ functions are independent of any derivative asset but generally dependent upon $(t, S_t, v_t)$. The Q-evolution model then becomes

$$dS_t = (r_t - q_t)S_t dt + \sigma_t S_t dW_{Q,t}^S,$$
$$dv_t = \beta_t^Q dt + \xi v_t dW_{Q,t}^v,$$

where $\lambda_t^e = (\alpha_t^P - r_t)/\sigma_t$ and $\beta_t^Q = \beta_t^P - \xi v_t \lambda_t^v$. Fixing $\beta_t^Q = \omega^Q - \kappa^Q v_t$ (where $\omega^Q \equiv \kappa^Q \bar{v}^Q$) from our postulated Q-measure GARCH diffusion (1) still leaves a *lot* of freedom for the P-evolution. In this generality, the only remaining compatibility requirements under "no-arbitrage" are:

- $\alpha^P(v_t = 0) = r_t$, since a stock holding would be instantaneously riskless, presuming a deterministic short-rate $r_t$.
- the boundaries $v = 0$ and $v = \infty$ should be unattainable in finite time by the P-measure $v_t$-process, since that is true of the postulated Q-measure process.

However, the spirit of the model is that the P-measure evolution is *also* a GARCH diffusion (recall the origin of the name). So, let $\beta_t^P = \omega^P - \kappa^P v_t$, with possibly different P-parameters. For example, let's postulate i) a volatility-dependent equity risk premium $\alpha_t^P = r_t + c v_t$, where $c$ is a (positive) constant, and ii) $\lambda^v$ is also constant. With those choices, our associated P-model GARCH diffusion is

$$dS_t = (r_t - q_t + c\, v_t) S_t dt + \sigma_t\, S_t dW_{P,t}^S,$$
$$dv_t = (\omega^P - \kappa^P v_t)\, dt + \xi v_t\, dW_{P,t}^v,$$

where $\omega^Q = \omega^P$ and $\kappa^Q = \kappa^P + \lambda^v \xi$. Now two additional parameters $(c, \lambda^v)$ need to be estimated and "P/Q compatibility" under our choices becomes a hypothesis to be tested. All that is outside our scope in this article. However, one expects $c > 0$ and $\lambda^v < 0$ for, say, SPX.

Finally, note that with our choices and using $(x_t, v_t)$ where $x_t = \log S_t$, *both* real-world and risk-neutral processes are MAPs (Markov Additive Processes). As discussed in the body, the MAP property leads to an "all-options-at-once" KBE solution for vanilla options; in addition, it also allows dimensional reduction by Fourier methods.

# Appendix B – Critical points for the stand-alone volatility process

In our model, the stand-alone volatility process evolves as $dV_t = \kappa(\bar{v} - V_t) dt + \xi V_t dB_t$, where $B_t$ is a Brownian motion. The same functional form holds under either measure (P/Q), although the numerical values of the parameters may differ. Let $p(t, v, v_0)$ denote the transition probability density for the process; i.e. $p(t, v, v_0) dv \equiv \Pr(V_t \in dv | V_0 = v_0)$. Then, $v_q = v_q(q, t, v_0, \kappa, \bar{v}, \xi)$, the $q$-critical point for the associated distribution, is defined by $\int_0^{v_q} p(t, x, v_0)\, dx = q$, where $0 \leq q \leq 1$. Having $v_q$ (for $q$ close to 1) is useful for setting the $v$-grid (upper) truncation points for the full 2D process PDE solvers.

Now $p(t, v, v_0)$ is not known analytically, but solves the Fokker-Planck problem:

$$\frac{\partial p}{\partial t} = -\frac{\partial J}{\partial v}, \quad \text{where} \quad J(t, v) \overset{\text{def}}{=} -\frac{1}{2} \xi^2 \frac{\partial}{\partial v} \{v^2 p\} + \kappa(\bar{v} - v) p, \qquad v \in (0, \infty),$$

and subject to the initial condition $p(0, v, v_0) = \delta(v - v_0)$, using the Dirac-delta. Here $J(t, v)$ is the probability current or flux. Mathematically, since both $v = 0$ and $v = \infty$ are inaccessible to the process (with $v_0 > 0$), no boundary conditions are necessary in the continuum problem.

**Two relations.** It is easy to find that in the limit $t \to \infty$, $v_t$ follows an Inverse Gamma distribution $p(x) = \beta^\alpha x^{-\alpha-1} e^{-\beta/x} / \Gamma(\alpha)$, with shape parameter $\alpha = 1 + 2\kappa/\xi^2$ and scale parameter $\beta = 2\kappa\bar{v}/\xi^2$. Another easy relation for arbitrary $t$ is the scaling identity:

$$v_q(q, t, v_0, \kappa, \bar{v}, \xi) = \bar{v} \times v_q\left(q, \xi^2 t, \frac{v_0}{\bar{v}}, \frac{\kappa}{\xi^2}, 1, 1\right),$$

which reduces the effective number of parameters by two. While the scaling relation was not used in the implementations, it was checked.

**Mathematica implementation.** When speed is not a factor, this full problem (solving the Fokker-Planck PDE, and calculating the critical point) is readily solved in Mathematica. Our short implementation is shown in Fig. 6. Even if you are not a Mathematica user, the syntax should be largely readable. The basic idea is to convert to new coordinates $x = \log v$ and solve the resulting PDE problem using NDSolve. A uniformly-spaced $x$-grid with $2 N_x$ points is centered at $x_0 = \log v_0$. The grid is truncated at $\pm n_1$ "sigma's" from $x_0$, where one sigma equals $\xi\sqrt{T}$. Numerical boundary conditions are taken to be (i) zero flux at $x_{min}$ and (ii) a zero spatial derivative at $x_{max}$. The initial condition is a lattice Dirac-delta, non-zero only at $x_0$, which lies exactly on a node.

**C/C++ implementation.** The Mathematica implementation solves the above Fokker-Planck problem using $4^{th}$ order spatial discretization and the Method of Lines (MOL) via an ODE solver in time. This yields very accurate results but is slow (though we have not tried to port it to C++). For the tests presented in this paper we have opted for a more standard approach which we only briefly outline here:

The discretization is based on uniform central second order finite differences and the Crank-Nicolson scheme with Rannacher time-stepping. Boundary conditions are the same as above. This approach works, apart from the far-left region of the grid where convection may dominate and result in oscillations/ negative densities. In such cases we locally introduce the 1st order upwind scheme for the convection term. If negative densities are still produced, we try to bump up $N_x$. If all fails, we simply return $v_q$ from the stationary (Inverse Gamma) distribution. We also apply double spatial Richardson extrapolation (which should in theory result in 6th order accuracy- if the upwind scheme is only used in areas where the density takes negligible values). All this results in accuracy even higher than Mathematica's, requiring CPU times of about 2-3 milliseconds with $N_x = 800$ and $N_T = 50$.

```mathematica
CriticalPointGarchPDE[q_, V0_, T_, NX_, vbar_, kappa_, xi_, n1_, AG_] :=
  Module[{X0, Xmin, Xmax, mu = kappa + 0.5 xi^2, omega, A,
      h0, h, dX, i, grida, gridb, grid, t, soln, cdf, critx, critv, critx99},
  
  Off[NDSolve::eerri];
  X0 = N[Log[V0]];
  Xmin = X0 - n1 xi Sqrt[T];
  Xmax = X0 + n1 xi Sqrt[T];
  dX = (X0 - Xmin) / NX;
  omega = kappa vbar;
  grida = Xmin + N[Table[i dX, {i, 0, NX}]];
  grida[[NX + 1]] = X0;
  gridb = X0 + N[Table[i dX, {i, 1, NX}]];
  grid = Join[grida, gridb];
  h0[X_ ?NumericQ] := If[Abs[X - X0] < 0.5 dX, 1 / dX, 0];
  Clear[soln, h, t]; Off[NDSolve::eerr];
  
  (* using zero flux condition at Xmin *)
  soln = h /. NDSolve[{∂_t h[x, t] == 0.5 xi^2 ∂_{x,x} h[x, t] - ∂_x ((omega E^(-x) - mu) h[x, t]),
      h[x, 0] == h0[x], 0.5 xi^2 ((D[h[x, t], x]) /. x → Xmin) - ((omega E^(-Xmin) - mu) h[Xmin, t]) == 0,
      ((D[h[x, t], x]) /. x → Xmax) == 0}, {h}, {x, Xmin, Xmax}, {t, T, T}, AccuracyGoal → AG,
      Method → {"MethodOfLines",
        "SpatialDiscretization" → {"TensorProductGrid", "Coordinates" → grid}}][[1]];
  
  A = NIntegrate[soln[x, T], {x, Xmin, X0}, AccuracyGoal → AG];
  cdf[y_ ?NumericQ] := A + NIntegrate[soln[x, T], {x, X0, y}, AccuracyGoal → AG];
  critx99 = y /. FindRoot[cdf[y] == 0.99, {y, X0, Xmin, Xmax}];
  critx = y /. FindRoot[cdf[y] == q, {y, critx99, Xmin, Xmax}];
  critv = E^critx;
  Return[critv]]
```

**Figure 6.** Mathematica code computing the stand-alone V-distribution critical points.

# Appendix C – Data description

**Option Chain A**. End-of-day (EOD) SPX option data on March 31, 2017 was obtained from the CBOE's LiveVol service: "End-of-Day Option Quotes with Calcs". These files record option quotes and CBOE calculated option implied volatilities (IV's) at 15:45 New York time. This time is 15 minutes prior to the regular session close in both NYC and Chicago. From the CBOE (edited for brevity):

"Implied volatility and Greeks are calculated off the 1545 timestamp, considered a more accurate snapshot of market liquidity than the end of day market. LiveVol applies a unified calculation methodology across both live and historical data sets to provide maximum consistency between back-testing and real-time applications. Cost of carry inputs (interest rates, dividends) are determined by a statistical regression process .... The cost of carry projected from these inputs is compared against those implied by the at-the-money options from each option expiry. If the rates differ significantly—and the option spreads for this expiry are sufficiently narrow—the implied rates replace the standard inputs. ...".

**Table 9.** SPX Option Implied Volatilities (%): Chain A (15:45, March 31, 2017).

| | Option expiration date (mm/dd/yy format) | | | | | | | |
|---|---|---|---|---|---|---|---|---|
| Strike | 4/21/17 | 5/19/17 | 6/16/17 | 9/15/17 | 12/15/17 | 6/15/18 | 12/21/18 | 12/20/19 |
| 500 | | | | | | 52.26 | 41.57 | 35.38 |
| 600 | | | | | 52.15 | 46.78 | 40.07 | |
| 700 | | | | | 51.44 | 42.40 | 37.49 | 33.29 |
| 800 | | | | 54.23 | 45.09 | 40.27 | 36.22 | 31.71 |
| 900 | | | | 48.05 | 41.67 | 36.41 | 33.23 | 30.24 |
| 1000 | | | | 45.31 | 38.94 | 34.48 | 32.19 | 29.50 |
| 1100 | | | 54.84 | 40.55 | 36.63 | 32.52 | 29.90 | 27.83 |
| 1200 | | | 49.69 | 36.68 | 34.26 | 30.83 | 29.05 | 26.78 |
| 1300 | | | 44.14 | 34.62 | 32.09 | 29.12 | 27.27 | 25.78 |
| 1400 | | | 41.26 | 32.49 | 29.90 | 27.51 | 25.94 | 24.76 |
| 1500 | | 45.17 | 37.60 | 30.19 | 27.89 | 25.96 | 24.80 | 23.80 |
| 1600 | | 40.28 | 33.48 | 27.92 | 26.02 | 24.58 | 23.54 | 22.91 |
| 1700 | | 36.19 | 30.91 | 25.78 | 24.23 | 23.13 | 22.38 | 21.91 |
| 1800 | 41.00 | 32.52 | 27.66 | 23.62 | 22.37 | 21.78 | 21.28 | 21.09 |
| 1900 | 34.65 | 27.88 | 24.43 | 21.56 | 20.78 | 20.44 | 20.15 | 20.14 |
| 2000 | 30.73 | 23.75 | 21.39 | 19.57 | 19.07 | 19.09 | 19.14 | 19.42 |
| 2100 | 24.74 | 19.72 | 18.34 | 17.55 | 17.50 | 17.78 | 18.06 | 18.59 |
| 2200 | 17.67 | 15.80 | 15.28 | 15.43 | 15.76 | 16.47 | 16.95 | 17.79 |
| 2300 | 11.61 | 12.13 | 12.40 | 13.36 | 14.06 | 15.20 | 15.93 | 16.89 |
| 2375 | 7.86 | 9.65 | 10.38 | 11.82 | 12.82 | 14.32 | 15.25 | 16.51 |
| 2400 | 7.37 | 9.02 | 9.80 | 11.35 | 12.43 | 14.01 | 14.97 | 16.34 |
| 2425 | 7.34 | 8.52 | 9.27 | 10.90 | 12.04 | 13.70 | 14.74 | 16.12 |
| 2450 | 7.95 | 8.15 | 8.91 | 10.46 | 11.65 | 13.38 | 14.43 | 15.99 |
| 2475 | 8.96 | 8.31 | 8.68 | 10.05 | 11.27 | 13.09 | 14.29 | 15.83 |
| 2500 | 10.43 | 8.18 | 8.66 | 9.71 | 10.92 | 12.80 | 13.99 | 15.66 |
| 2525 | 12.20 | 8.77 | 8.78 | 9.46 | 10.61 | 12.50 | 13.76 | 15.50 |
| 2550 | 12.68 | 9.46 | 8.92 | 9.25 | 10.29 | 12.25 | 13.56 | 15.36 |
| 2575 | 13.53 | 10.14 | 9.20 | 9.11 | 10.05 | 11.98 | 13.35 | 15.18 |
| 2600 | 14.49 | 11.02 | 9.54 | 9.06 | 9.86 | 11.72 | 13.14 | 15.01 |
| 2625 | | | 10.01 | 9.07 | 9.65 | 11.41 | 12.94 | 14.85 |
| 2650 | | | 10.53 | 9.11 | 9.53 | 11.23 | 12.85 | 14.75 |
| 2675 | | | | | 9.43 | | 12.54 | |
| 2700 | | | 11.27 | 9.35 | 9.40 | 10.79 | 12.37 | 14.39 |
| 2750 | | | | 9.72 | 9.37 | 10.45 | 12.05 | 14.09 |
| 2800 | | | | 10.08 | 9.40 | 10.17 | 11.74 | 13.81 |
| 2850 | | | | | | 9.99 | | 13.55 |
| 2900 | | | | | 9.74 | 9.91 | 11.17 | 13.32 |
| 3000 | | | | | 10.79 | 10.04 | 10.79 | 12.87 |
| 3100 | | | | | | | 10.66 | 12.46 |
| 3200 | | | | | | | 10.63 | 12.18 |
| 3300 | | | | | | | | 11.95 |
| 3400 | | | | | | | 11.86 | 11.45 |
| 3500 | | | | | | | 11.75 | |

On any given data, option data can be quite voluminous and needs to be filtered, both to reduce calculational burdens and remove irrelevant noise. Indeed, the full March 31, 2017 data file contained 8399 option line items, which we filtered to the 246 items shown in Table 9. This was done by, first, focusing on traditional "third Friday" expirations and then doing some strike filtering. The 15:45 SPX index value was $S_0 = 2367.94$ (midpoint quote). We selected positive bid, out-of-the-money options, so the IV's shown in the table are from puts when the strike $K < S_0$ and otherwise calls. (The CBOE's IV methodology is somewhat of a black-box, but it appears to be essentially put-call-parity preserving. See also [18]). For the first expiration we chose (Apr 21, 2017), the implied volatilities were smooth down to a strike $K = 1800$, chosen as a lower limit strike cutoff. For other expirations, the data looked smooth down to $K = 500$, our cutoff for the remaining expirations. We imposed no upper strike cutoff. To achieve a rough balance between puts and calls, we selected put strikes at multiples of 100 and call strikes at multiples of 25. We believe this filtering retains the important characteristics of the full data sets.

For short-term interest rates, we found U.S. Treasury debt asked yields on Mar 31, 2017 from the Wall Street Journal (WSJ) and use those as stepwise constants for each of our 8 expirations. That series was {0.00728, 0.00723, 0.00716, 0.00865, 0.00939, 0.01118, 0.01203, 0.01434} in expiration order. For the SPX dividend yield, we used a constant $q = 0.0197$ for all expirations, the WSJ-reported trailing 12-month SPX yield on the same date. These are not necessarily the cost-of-carry parameters used by the CBOE, the latter unavailable. The difference is unlikely to change the parameter fits or our conclusions in any way that matters. But if the reader is concerned about this small point, then take our combination of IV's and cost-of-carry's as one 'possible' market data set (largely consistent with the 3/31/17 actual data) to which we fit various models.

**Option Chain B.** The second author (Lewis) collected (out-of-the-money) closing option quotes at the time (Feb. 1, 2010), using only options with positive bids. IV's were calculated from the bid-ask midpoint option price. Interest rates and a dividend yield was found from the WSJ as per Chain A.

## Appendix D – Calibration for the randomized optimal $p$-model

A randomized version of the Heston model was presented by Mechkov [16]. The basic idea is that instead of taking the initial value of the (latent) variance process $v_0$ to be a known fixed value, we assume it is given by some distribution. It is a simple and appealing idea making for an easy extension to the present framework: The PDE solution automatically provides the option values for the whole range of possible initial variance values (corresponding to the grid in the $v$-direction). To find the "randomized" option price at the asset spot $S_0$, average the solution across the $S = S_0$ grid line using the assumed distribution. As suggested in [16], it is reasonable to assume the latter should be of the same type as that of the process' equilibrium distribution. For the GARCH diffusion model this would be the Inverse Gamma distribution. Our brief testing indicates that this choice yields the best results even for the randomization of the Heston model[32], so we will use it here to randomize the general $p$-model (32). We make the parameters of the distribution (shape $\alpha$ and scale $\beta$), as well as the power $p$ of the model part of the calibration. The total number of parameters to be fitted is now seven.

As Figure 7 shows, the overall quality of fit ($\text{RMSE}_{IV} = 0.79\%$) is considerably better than either that of the GARCH diffusion ($\text{RMSE}_{IV} = 1.68\%$) or the Heston model ($\text{RMSE}_{IV} = 1.28\%$), especially for the shortest expiration (see Figure 3). As already discussed in Sec. 4.5, the Heston model ($p = 0.5$) fit implies unlikely dynamics. The optimal power of the general model was calibrated to $p = 0.8$, slightly closer to the GARCH diffusion model[33]. In contrast to the latter, here we see a steep smile captured for the first (3W) expiration, which is due to the randomization (and not the change from $p = 1$ to $p = 0.8$). We also note how now that the model is not bound (and thus stressed) by its inability to account for the short-term smile, the fitted volatility of volatility parameter falls to arguably more realistic levels ($\xi = 2$). This is much lower than GARCH diffusion's calibrated value of $\xi = 6.9$ (Table 1), even after accounting for the scaling from $p = 1$ to $p = 0.8$.

A potential problem is seen with the calibrated correlation coefficient ($\rho = -1$). Other similar tests also indicate that the randomization procedure tends to lead the optimizer to rather extreme correlation values. Why? From [17], the small-$T$ asymptotic smile "explosion", due to randomization, is symmetric about $x_K = 0$, where $x_K = \log K/S_0$ is the log-moneyness. The optimizer may be trying to compensate for the 'unwanted' new tendency towards symmetry by pushing $\rho$ towards -1. Excluding $\rho$ from the calibration and fixing it to a more reasonable value ($\rho = -0.8$) yields a very similar fit with $\text{RMSE}_{\text{IV}} = 0.83\%$. This indicates that the much-improved fit does not depend strongly on such extreme $\rho$ values. Nevertheless, this behavior (as well as needing a dynamical rationale) are issues for randomization.

---

[32] The equilibrium (stationary) distribution of the square root variance process is a Gamma distribution, but at least for our datasets we found it actually performs worse than the Inverse Gamma as the randomizing (initialization) distribution for the Heston model.

[33] The model is much closer to the GARCH diffusion model in terms of the implied dynamics. For example, one finds that the Heston model's fit here implies a 41% probability that the long-run (risk-neutral) volatility is less than 1%, which is not very plausible. For both GARCH diffusion and the $p = 0.8$ model this probability is practically zero. The mean long-run volatility is about 20% for all three models.

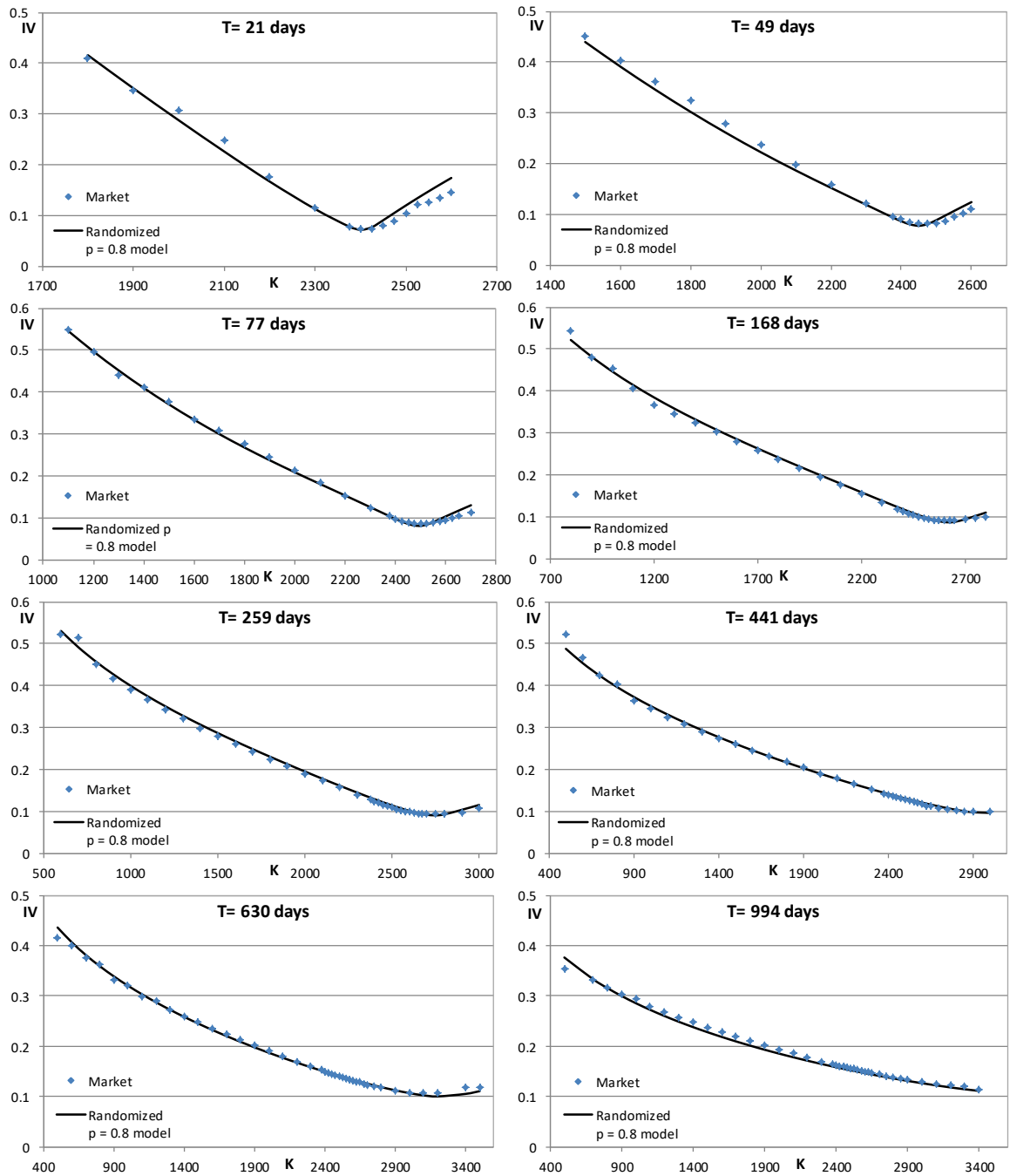

**Figure 7**. Calibration of the general power-law model (32) for Chain A, randomizing $v_0$ with an Inverse Gamma distribution. The calibrated model parameters are $\bar{v} = 0.0407$, $\kappa = 3.34$, $\xi = 2.00$, $\rho = -1.00$, the initial distribution shape and scale parameters (see Appendix B) $\alpha = 1.05$, $\beta = 0.00124$ and the model power $p = 0.801$. $\text{RMSE}_{\text{IV}} = 0.79\%$.